\documentclass[3p]{elsarticle}
  \makeatletter
    \def\ps@pprintTitle{%
      \let\@oddhead\@empty
      \let\@evenhead\@empty
      \let\@oddfoot\@empty
      \let\@evenfoot\@oddfoot
    }
    \makeatother
\journal{Journal of Sound and Vibration}

\usepackage{amsmath,amsfonts,mathtools}    % need for subequations
\usepackage{graphicx}   % need for figures
\usepackage{color}      % use if color is used in text
\usepackage{float}
\usepackage{fullpage}
\usepackage{subfig}  % use for side-by-side figures
\DeclarePairedDelimiter{\floor}{\lfloor}{\rfloor}

\renewcommand{\appendix}{
  \setcounter{section}{0}\renewcommand{\thesection}{\Alph{section}}
  \section*{Appendix} }

\def\Appendix#1{
  \setcounter{equation}{0}
  \renewcommand{\theequation}{\thesection\arabic{equation}}
  \section{#1}  }

\def\beq#1#2\eeq{\begin{equation}\label{#1}#2\end{equation}}
\def\bal#1#2\eal{\begin{align}\label{#1}#2\end{align}}
\def\bse#1#2\ese{\begin{subequations}\label{#1}#2\end{subequations}}
\def\ba{\begin{aligned}}
\def\ea{\end{aligned}}
\newcommand{\beqa}[1]{ \begin{eqnarray} \label{#1} }	
\newcommand{\eeqa}{\end{eqnarray}  }

\def\rev#1{{#1}} 
\begin{document}
\begin{frontmatter}
\title{ Acoustic scattering from an infinitely long cylindrical shell 
\\
with an internal mass attached by multiple axisymmetrically distributed stiffeners }

\author[add]{Alexey S. Titovich\corref{cor1}\fnref{fn1}}
\ead{alexey.titovich@rutgers.edu}

\author[add]{Andrew N. Norris}
\ead{norris@rutgers.edu}

\cortext[cor1]{Corresponding author}
\fntext[fn1]{Tel:1-848-445-2248}

\address[add]{Mechanical and Aerospace Engineering, Rutgers University, 98 Brett Road, Piscataway NJ 08854}

\date{\today}% It is always \today, today,  but any date may be explicitly specified

\begin{abstract}
\rev{A thin infinitely long elastic shell is stiffened by $J$ in number identical lengthwise ribs distributed uniformly around  the circumference and joined to a rod in the center.} The 2D model of the substructure is a rigid central mass supported by $J$ axisymmetrically placed linear springs. \rev{The response of the shell-spring-mass system is quite different from a fluid filled shell or that of a solid cylinder due to the discrete number of contact points which couple the displacement of the shell at different locations.} Exterior  acoustic scattering  due to normal plane wave incidence is solved in closed form for arbitrary $J$. The scattering matrix associated with the normal mode solution displays a simple structure, composed of distinct sub-matrices which decouple the incident and scattered fields into $J$ families. The presence of a springs-mass substructure causes resonances which are shown to be related to the subsonic shell flexural waves, and an approximate analytic expression is derived for the quasi-flexural resonance frequencies. Numerical simulations indicate that the new solution for $J\ge 3$ springs results in a complicated scattering response for plane wave incidence. \rev{As the number of springs becomes large enough, the total scattering cross-section is asymptotically zero at low frequencies and slightly increased compared to the empty shell at moderate frequencies due to the added stiffness and mass.} It is also observed that the sensitivity to the angle of incidence diminishes as the number of springs is increased. \rev{This system can be tuned by selecting the shell thickness, spring stiffness and added mass to yield desired quasi-static effective properties making it a candidate element for graded index sonic crystals.}

%A thin infinitely long elastic shell contains a rod at it center connected to the shell by $J$ identical lengthwise ribs distributed uniformly around  the circumference. The 2D model of the substructure is a rigid central mass supported by $J$ axisymmetrically placed linear springs.  Exterior  acoustic scattering  due to normal plane wave incidence is solved in closed form for arbitrary $J$. The scattering matrix associated with the normal mode solution displays a simple structure, composed of distinct sub-matrices which decouple the incident and scattered fields into $J$ families. The presence of a springs-mass substructure causes resonances which are shown to be related to the subsonic shell flexural waves, and an approximate analytic expression is derived for the quasi-flexural resonance frequencies. Numerical simulations indicate that the new solution for $J\ge 3$ springs results in a complicated scattering response for plane wave incidence with a strong dependence on the number of springs. Likewise, it is observed that the sensitivity to the angle of incidence diminishes as the number of springs is increased. These results as well as the possibility of tuning these shell-springs-mass systems, either passively or actively, to yield desired quasi-static effective properties are relevant to the design of acoustic metamaterials. 

\end{abstract}
\begin{keyword}
Acoustics \sep  Scattering \sep Shell 
\end{keyword}
\end{frontmatter}

\section{Introduction} \label{intro}

The scattering of acoustic waves from an elastic cylindrical shell with internal structure is quite distinct from the response of a simple shell. Excitation of waves on shells arises from two general mechanisms: (i) phase matching to supersonic membrane-type waves~\cite{Scott88, Rumerman93, Norris94a}, or (ii) excitation at structural discontinuities. The latter can excite both supersonic longitudinal waves which then re-radiate into the fluid, and subsonic flexural waves which can persist for long times and over large propagation paths on the structure. Flexural waves are an important source of structural energy transfer, but they are not usually excited on a smooth metallic shell in contact with an exterior acoustic medium such as water because of their subsonic phase speed. The effect of structural discontinuities or constraints can be modeled as effective forces on an otherwise smooth shell, analyzed  in the original work of Bleich and Baron \cite{Bleich54}. 

Structural constraints can be separated into three fundamental types: concentrated, linear circumferential, and linear lengthwise. Attachment of a spring-mass system or a beam to the interior surface of the shell constitutes a concentrated constraint. The constraint inhibits or enhances the vibration of the shell through reflection/conversion of structural waves as well as through the resonant behavior of the substructure itself. Undersea vehicles are sometimes modeled as a shell with many spring-mass oscillators attached to the interior. Analysis of such "fuzzy structures" indicates  possible wave localization due to structural irregularity, which in turn suggests methods for controlling vibration/scattering \cite{Pierce95, Houston95, Strasberg96, Photiadis97, Bucaro98, Photiadis00}.

The other type of constraint, circumferential discontinuities, include examples such as rigid discs~\cite{Skelton91}, plates, rings~\cite{El-Raheb89}, ribs, bulkheads~\cite{Coupry84, Guo94, Bjarnason94, Photiadis94, Cuschieri95a, Cuschieri95b, Loftman99, Tran-Van-Nhieu01, Tran-Van-Nhieu02}, and any other frames thin in the axial direction. Bloch-Floquet waves and Bragg scattering effects appear for  oblique incidence
if the internals are placed periodically along the axis of the shell, ~\cite{Tran-Van-Nhieu01, Tran-Van-Nhieu02}. Analysis  of oblique incidence onto shells with several bulkheads show that constructive interference between the scattered pressure due to each bulkhead produces a dipole-like radiation pattern and scattered pressure associated with bending moments yields a quadrupole-like radiation pattern \cite{Cuschieri95a, Cuschieri95b}.

This paper is concerned with the two dimensional (2D) modeling of lengthwise sheet springs supporting an internal mass as explored in~\cite{Achenbach92, Guo92, Ho96}, which can be viewed as lengthwise discontinuities. More sophisticated and certainly more realistic models such as deck-type plates~\cite{Bjarnason92, Klauson92, Guo93, Guo96, Baillard00} and lengthwise elastic ribs~\cite{Klauson92, Klauson94} also fall into the category of lengthwise discontinuities. These internal structures provide more mechanisms for coupling to and mixing of the structure-borne waves producing a very complex response. With normal wave incidence and a sufficiently long shell, lengthwise constraints can be analyzed in two dimensions as will be done herein.

Understanding of cause and effect can be obtained through detailed analysis of simple models for internal structure. The simplest model for internal structure is a single mass attached by a single spring to the shell. The structural analogue of this system is a long internal rod attached to the shell by a lengthwise rib. Although springs cannot support the passage of waves, this is a rich and relatively complex system as compared to the bare shell, and it displays many of the dynamic properties of much more complex substructures. The first such analysis by Achenbach et al.\ \cite{Achenbach92} considered the 2D problem of a shell with an internal mass supported by a single spring and loaded by an external point force. Via an energy formulation the interaction force between the spring-mass system and the shell was determined and its affect on the acoustic scattering studied, especially in the vicinity of the spring-mass resonance. The presence of the substructure generates acoustic radiation which can be greater or lesser than that of the standalone shell  based on the frequency of the harmonic excitation relative to the resonance of the oscillator (spring-mass system).

The problem becomes more complicated when the mass is supported by more than one spring. Guo \cite{Guo92} formulated the scattering solution for a shell with an internal mass attached by a diametrical pair of springs (structural analogue being a rod supported by a diametric pair of lengthwise ribs). He demonstrated that there are two distinct solutions, for even and odd azimuthal modes, which superimpose to produce the overall response of the shell-springs-mass system. This simple model clearly reveals the rich and complex set of resonances resulting from flexural waves excited by the spring attachments. This stiffener-borne wave generation mechanism was investigated earlier by Klauson and Metsaveer \cite{Klauson92}. Guo showed that the addition of a dissipative mechanism into the springs-mass system did little to the scattered field. Later, Gaunaurd \cite{Gaunaurd93} expanded the analysis by considering a neutrally buoyant spherical shell with a double spring-mass system. Spectral theory was used by Ho \cite{Ho96} to obtain the acoustic response for a shell with the mass supported by a non-diametrical pair of springs.

The current work reconsiders \rev{acoustic scattering at low to mid-frequencies, $ka\le20$, from a shell  with} simpler internal structural models, focusing on a distribution of an arbitrary number of $J$  springs supporting a central internal mass. This is an approximate 2D model of a central rod supported by an equally spaced distribution of $J$ lengthwise ribs. \rev{The shell-spring-mass system is particularly interesting because of how differently it responds to an incident wave when compared to fluid filled shells or solid cylinders. Primary reasons for studying such systems include understanding: 1) the acoustic scattering from a shell with a finite number of coupled point forces along the circumference, 2) the propagation of flexural-borne waves into the far-field for different number of springs, 3) the shift in resonant frequencies of the flexural waves due to the added stiffness, 4) low-frequency transparency with large number of springs and 5) the effect of the angle of incidence on scattering. Furthermore, the acoustic response of the shell changes by selecting different spring stiffness and added mass. This ability to tune the shell expands the range of possible acoustic properties for shells presented in Martin et al.~\cite{Martin12} and thus makes it a perfect element in graded index sonic crystals. Those results will be presented in a forthcoming paper. Here we concentrate on deriving and quantifying the model for arbitrary number of internal springs. } 

The  model considered expands the existing results \cite{Achenbach92, Guo92} for masses attached by one or two springs, to the more general case of $J$ attachment springs, where $J\ge 1$ is arbitrary. For an axisymmetric distribution of such springs, we utilize the symmetry of the problem to simplify the interaction force, which is later used to determine the T-matrix of the combined system. The results are presented successively for $J=1$, $J=2$, and finally $J\ge 3$ springs. The T-matrix is expressed in terms of physical quantities: acoustic, shell and spring impedance. These combine in a non-trivial way by virtue of the problem formulation to give the total impedance of the combined system. This total impedance governs the system's resonant behavior.

The layout of the paper is as follows. We begin in \S\ref{sec1} with a definition of the problem and a summary of the main results. The governing equations for the shell and the acoustic medium are given in \S\ref{sec2}. The single mass and multiple spring attachment model is described in \S\ref{sec3}. The main results for scattering from a shell with this  internal substructure are presented in \S\ref{sec4}. Properties of the general solution are discussed in \S\ref{sec5}. It is shown  that the scattered field decomposes into $J$ distinct parts, and  that the additional portion of the T-matrix due to the internal springs-mass system can be expressed by $J$  products of vectors, convenient for implementation. Numerical examples are given in \S\ref{sec6} along with a discussion of the backscatter and total scattering cross-section for plane wave incidence and various spring distributions. Approximate but useful expressions are derived for the effective resonant frequencies of the shell-springs-mass system. Conclusions are presented in \S\ref{sec7}.

\section{Problem definition and summary of results} \label{sec1}

\subsection{Scattering formulation}

\begin{figure} [H]
\begin{center}
\includegraphics[width=3.in]{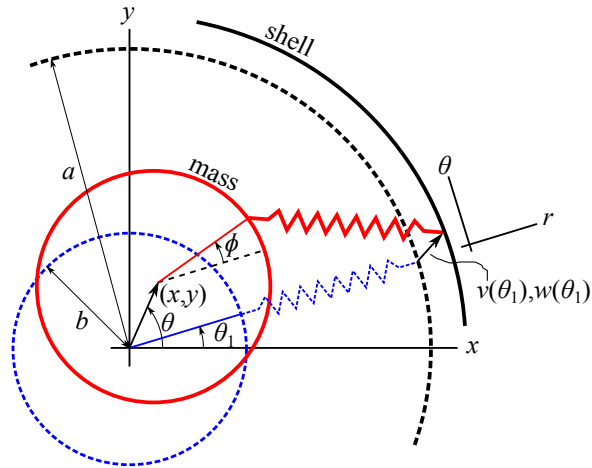}
\caption{Displaced shell and internal mass (shown with solid lines) connected by a single spring initially oriented radially at an angle $\theta_1$ from the x-axis.
} \label{fig2}
\end{center}
\end{figure}

Consider in-plane acoustic wave scattering from a thin cylindrical shell immersed in an acoustic  medium with volumetric mass density $\rho$ and sound speed $c$. A single mass per unit axial length $m$ is attached to the inner surface of the shell by a set of $J\ge 1$ springs each of stiffness $\kappa$ (with units of force per unit area) oriented at angles \rev{$\theta_j$ with respect to the horizontal, where $j=1,\dots,J$}. The springs are assumed equally distributed, so that  $\theta_{j+1} = \theta_j + 2\pi/J$. The mass is of finite size and free to rotate, as shown in the schematic in Fig.\ \ref{fig2}.
The thin shell has outer radius $a$, thickness $h\, (\ll a)$, volumetric mass density $\rho_s$, with elastic properties characterized by Young's modulus $E$ and Poisson's ratio $\nu$. We assume time dependence $\text{e}^{-\text{i} \omega t}$, which is henceforth omitted but understood. The total acoustic pressure on the shell $p$ satisfies the Helmholtz equation
\begin{equation} \label{10}
\nabla^2 p + k^2 p = 0,
\end{equation}
where $k=\omega/ c$ is the acoustic wavenumber. The pressure can be decomposed into two parts, the incident and scattered fields, $p_i$ and $p_s$ respectively, each a separate solution of Helmholtz's equation. Here we consider in-plane or 2D motion with plane wave incidence, requiring only the planar modes.  Thus,  
\begin{equation} \label{13}
p = p_i  + p_s,  \quad 
p_i = \sum\limits_{n = - \infty}^\infty  A_n J_n(kr)  \text{e}^{\text{i} n \theta} , \quad 
p_s = \sum\limits_{n = - \infty}^\infty  B_n H_n^{(1)}(kr) \text{e}^{\text{i} n \theta}, \quad r\ge a 
\end{equation}
with $A_n$ the incident field coefficients, $B_n$  the scattering coefficients, $J_n$ the Bessel function of the first kind of order $n$ and $H_n^{(1)}$ the Hankel function of the first kind of order $n$.

The objective is to get a relation between the incident amplitudes $A_n$ and the scattering amplitudes $B_n$. The solution is embodied in the infinite T-matrix defined by 
\beq{-41}
\bf B = \bf T \bf A  ,
\eeq
where $\bf A $ and $\bf B$ are vectors of infinite length comprised of the elements $A_n$ and $B_n$ at position $n\in\mathbb Z$, respectively.

\subsection{Summary of the main results}

\rev{As is customary in acoustics, the scattering solution is expressed in terms of various modal impedances, e.g. $Z_n$, $Z_n^{sh}$, $Z_n^{sp}$, each of which relates  radial stress to   radial velocity, see for example eq.\ \eqref{-5}.} Define the following impedances associated with azimuthal mode $n$,   
\begin{subequations}\label{7=8}
\begin{align}
Z_n &= \text{i} \rho c \frac{H_n^{(1)}(ka)}{H_n^{(1)\prime}(ka)}  ,
\quad 
\widehat Z_n = \text{i} \rho c \frac{J_n(ka)}{J_n^ \prime(ka)} ,
\label{3=3}
\\
Z_n^{sh} &=    - \text{i} \rho_s c_p \frac{ h}{a}
%\frac{  \big[ \Omega^4 - \Omega^2(1+n^2+\beta^2 n^4) + \beta^2 n^6 \big] } {\Omega  (\Omega^2-n^2)}  
 \Big[ \Omega- \frac{\beta^2 n^4}{\Omega}  - \Big( \Omega- \frac{n^2}{\Omega} \Big)^{-1}\Big] 
,  \label{02} 
\\ 
Z^{sp}_n(J) &= \frac{\text{i} J \kappa}{2 \pi a \omega } \times
\begin{cases}
\frac{1}{1 - H_J \frac{ \omega_{sp}^2 }{ \omega^2} } ,&   n= \pm 1 \,\text{mod}\, J,  
\\
1 ,&  \text{otherwise}, 
\end{cases}
\quad , \quad  
H_J = 
\begin{cases}
J , & J=1,2,
\\
\frac{J}{2}, & J \ge 3. 
\end{cases}
\end{align}
\end{subequations}
where $\omega_{sp}$ is a natural frequency of the springs-mass system and $\Omega$ a non-dimensional measure of frequency based on the shell extensional wavenumber , 
\beq{4=4}
\omega_{sp}^2=\frac{\kappa}m , \qquad
\Omega = \frac{\omega a}{c_p} \ \ \Big( = \frac c{c_p} ka\Big), 
\eeq
and the additional shell parameters are $c_{p}^{2} = {E}/[{\rho_s (1-\nu^2)}]$ and $\beta = \frac{1}{\sqrt{12}}\frac{h}{a}$. \rev{Different azimuthal modes are affected differently by the spring-mass system (see $Z^{sp}_n$) where the function $n= \pm 1 \,\text{mod}\, J$ is used in this paper to mean 
\begin{equation} \label{nmod}
n= \pm 1 \,\text{mod}\, J \quad \Longleftrightarrow \quad n=\pm 1+mJ, \,\, \text{where} \,\, m=0,\pm1, \pm2, \pm3, ...
\end{equation}
such that $J$ is the modulus of the congruence.} The various impedances can be interpreted as follows: $Z_n$ is a radial acoustic impedance associated with radiating wave functions, as compared with $\widehat Z_n$ for regular wave functions; $Z_n^{sh}$ is the shell impedance; and $Z^{sp}_n$ is a generalized spring impedance, see \S\ref{sec3}. The expression for $Z_n^{sh}$ is based on the Donnell-Mushtari thin shell model, see \S\ref{sec2.1}, which is sufficient for the range of frequencies considered ($ka\le20$), although other expressions could be used, including the exact result from elastodynamics. Regardless of the specific shell model, the results in eqs.\  \eqref{-46} and \eqref{8} retain their analytic structure. The total equivalent impedance $Z_n^{tot} $ is defined by the series/parallel combination of the above impedances as
\begin{equation} \label{-46}
\frac 1{Z_n^{tot}} 
=  \frac{1}{Z^{sp}_{n}}  +\sum\limits_{p = - \infty}^\infty \frac{1}{Z^{sh}_{n+pJ}+Z_{n+pJ}} .
\end{equation}

Our main result is that the T-matrix has the following form ($^*$ denotes the complex conjugate)
\bse{8}
\bal{8a}
{\bf T} &= {\bf T}^{(0)} +  \sum\limits_{j=1}^{J} {\bf b}_j {\bf b}_j^T   \quad \text{where}
\\
{\bf T}^{(0)} &= \text{diag}(T_n) , \quad \quad 
T_n = \frac 12\left( \frac{\zeta_n^*}{\zeta_n} - 1\right), 
\quad
\zeta_n = (Z_n^{sh} + Z_n) H_n^{(1)\prime}(ka) ,
\quad 
\label{8b}
\\
b_{j,n} &= \frac{\text{i}}{\zeta_n} % (Z_{n}^{sh}+Z_{n})H_{n}^{(1)\prime}(ka)} 
\left(  \frac{2\rho c Z^{tot}_n}{\pi ka}\right)^{1/2}  \ \ \text{if}\ 
n=j\,\text{mod}\, J  , \  \text{otherwise}\ 0 . 
\label{8c}
\eal
\ese
These results are derived next. 

\section{Elastic shell in fluid} \label{sec2}
\subsection{Shell and acoustic pressure equations}\label{sec2.1}
The equations of motion for a thin cylindrical shell in the $r$ and $\theta$ directions, respectively, are \cite{Junger86}
\begin{subequations} \label{eom}
\begin{align}\label{eoma}
\frac{1}{a^2} \frac{\partial v}{\partial \theta} + \frac{w}{a^2} + \frac{\beta^2}{a^2} \frac{\partial^4 w}{\partial \theta^4}  + \frac{\ddot{w}}{c_{p}^{2}}  &=  \frac{\sigma(\theta,t)}{\rho_s c_p^2 h} ,
\\
\frac{1}{a^2} \frac{\partial^2 v}{\partial \theta^2} + \frac{1}{a^2} \frac{\partial w}{\partial \theta} - \frac{\ddot{v}}{c_{p}^{2}} &= 0 ,
\label{eomb}
\end{align}
\end{subequations}
where $w$ and $v$ are the radial and azimuthal displacement, respectively, $\sigma$ is the normal stress acting in the radial direction. The displacements and the forcing take the form
\begin{equation} \label{disp}
(w,\, v,\, \sigma)  = \sum\limits_{n=-\infty}^{\infty}\big( W_n,\, V_n,\, \sigma_n \big)\,  \text{e}^{\text{i} n \theta} . 
\end{equation}
Substituting equations \eqref{disp} into \eqref{eom} gives the modal equations as
\beq{modal}
\begin{aligned} 
(-\Omega^2 + 1 + \beta^2 n^4) W_n + \text{i} n V_n &=   \frac{a^2 \sigma_n}{\rho_s c_p^2 h}  ,
\\
\text{i} n W_n + (\Omega^2 - n^2)V_n  &= 0 .
\end{aligned}
\eeq
In the presence of forcing $\sigma_n$ the radial displacement may be defined in terms of a shell impedance $Z_n^{sh}$, 
see \eqref{02}, as
\beq{-5}
\sigma_n = -\text{i} \omega Z_n^{sh} W_n .\eeq
Note that the shell impedance $Z_n^{sh}$ is either mass or stiffness-like, depending on the frequency. 
The natural frequencies of the shell correspond to the  existence of nontrivial solutions in the absence of loading, and hence are defined as the roots of  
$Z_n^{sh} (\Omega ) = 0$.

Continuity between the radial shell velocity and the radial particle velocity in the fluid, combined with the momentum equation in the fluid implies, using eq.\  \eqref{13}, that 
$\ddot{w}  = - \rho^{-1}{\partial p }/{\partial r}$ on $r=a$, hence
\beq{-221}
%\ddot{w}  = - \frac 1{\rho}\frac{\partial p }{\partial r}  \ \ \text{on} \ r=a ,\quad \Rightarrow \quad
\rho c \, \omega \, W_n = A_n   J_n^\prime(ka) + B_n   H_n^{(1)\prime}(ka) . 
\eeq
Expanding the surface pressure as
\beq{-4}
p(a,\theta) = \sum\limits_{n=-\infty}^{\infty}P_n\,  \text{e}^{\text{i} n \theta}
\eeq
yields the coefficients for the scattered pressure and for  the total pressure  on the shell surface in terms of the radial displacement 
\bse{2-1}
\bal{Bn}
B_n &= \frac{1}{H_n^{(1)\prime}(ka)} \big[ \rho c \omega W_n  -  J_n^\prime(ka) A_n\big]  ,  
\\
P_n &=  -\text{i} \omega Z_n W_n   + \frac{2 \text{i}}{\pi ka} \frac{A_n}{H_n^{(1)\prime}(ka)} ,
\label{2-2}
\eal
\ese
where we have used the Wronskian identity $J_n(x) H_n^{(1)\prime}(x) - J_n^\prime(x) H_n^{(1)}(x) = \frac{2\text{i}}{\pi x} $. Equations \eqref{2-1} are valid whether or not the internal substructure is present. 

\subsection{Scattering in the absence of internal substructure}\label{submerged}

With no substructure inside the shell, the radial forcing on the shell is simply that of the incident and scattered pressure: $\sigma = -p$. 
Therefore, combining \eqref{-5}, \eqref{2-2} with $\sigma_n =-P_n$ 
and the definition of $\zeta_n$ in \eqref{8b} gives $W_n = 2 A_n / (\pi \omega ka \zeta_n ) $. Equation \eqref{Bn}  then yields $B_ n = T_n A_n$ where the (diagonal) T-matrix elements $T_n$ are defined in eq.\ \eqref{8b}.  
The associated element of the diagonal "S-matrix"  is  
\beq{-23}
S_n = 1+ 2 T_n  \implies 
S_n= \text{e}^{-\text{i} 2\phi_n} ,  \ \ 
T_n  = -\text{i} \text{e}^{-\text{i} \phi_n} \sin\phi_n,  \ \ 
\text{with} \ 
\phi_n = \text{arg} \,\zeta_n,
\eeq
implying that $|S_n|=1$, $|T_n|\le 1$, in conformity with the fact that no dissipation is assumed.

\section{The springs-mass model}\label{sec3}

Consider now the mass per unit length, $m$, attached to the shell as shown in Fig.\ \ref{fig2} by $J\ge 1$ springs each of stiffness per unit area, $\kappa$, oriented at angles \rev{$\theta_j$, where $j=1,\ldots,J$}. The springs are assumed equally distributed, so that $\theta_{j+1} = \theta_j + 2\pi/J$. The horizontal and vertical displacements of the mass are denoted by $x$ and $y$, respectively. The derivation of the linearized equations of motion for the internal mass and the resulting radial force on the shell are in Appendix \ref{appendixA}. In summary, the displacement of the finite sized mass associated with its rotation is of second order and not retained in the linearized equations. Moreover, the angular motion of the mass is not excited by the acoustic incidence. Only the translating degrees of freedom of the mass contribute to the radial force on the shell. Introduce the force distribution per unit area of the shell surface, $f(\theta)$, defined such that $f d A $ is the force acting on an element $dA =a d \theta \, dz$. It follows from the Appendix that  in the case of one, two, and $J\ge 3$ springs, respectively,
\begin{subequations}\label{8080}
\begin{align}
f(\theta) =& - \frac{\kappa}{a} \left( \frac{\tau^2}{\tau^2 - 1} \right) w(\theta_1) \delta(\theta-\theta_1)  , \ \ \text{one spring},
\\
f(\theta) =& -\frac{\kappa}{a} \left( \frac{1}{\tau^2 - 2} \right) \bigg[ \Big( (\tau^2-1)w(\theta_1) - w(\theta_2) \Big)  \delta(\theta-\theta_1) 
\notag \\
&+ \Big(  (\tau^2-1)w(\theta_2) - w(\theta_1) \Big) \delta(\theta-\theta_2)  \bigg]  ,
\ \ \text{two springs}, \ \ (\theta_2 = \theta_1 +\pi)
\\
f(\theta) =& -\frac{\kappa}{a} \left( \frac{1}{\tau^2 - \frac{J}{2}} \right) \sum\limits_{j=1}^{J} \bigg[ \sum\limits_{n=1}^{J} w(\theta_n) \cos(\theta_j-\theta_n) + \Big(\tau^2 - \frac{J}{2}\Big) w(\theta_j) \bigg] \delta(\theta-\theta_j) ,
\end{align}
\end{subequations}
\rev{where $\delta(\theta)$ is the Dirac delta function and (see eq.~\eqref{4=4}) }  
\beq{3333}
\tau^2 = \frac{\omega^2}{\omega_{sp}^2} \ \ \Big( =\frac{m \omega^2}{\kappa} \Big).
\eeq
Expanding the radial force distribution of eq.\ \eqref{8080} in azimuthal modes as
\begin{equation} \label{-19}
f(\theta) = \sum\limits_{n=-\infty}^\infty f_n \text{e}^{\text{i} n \theta} ,  
\end{equation}
and using the identity $\delta(\theta-\theta_j) 
= \frac 1{2\pi} \sum\limits_{n= -\infty}^\infty
\text{e}^{\text{i} n (\theta  -\theta_j) }
$ 
gives the modal force on the shell for the cases of one, two, and $J\ge 3$ springs as
\begin{subequations} \label{_25}
\begin{align}
f_n &= - \frac{\kappa}{2\pi a} \Big(\frac{\tau^2}{\tau^2 - 1} \Big) w(\theta_1) \text{e}^{-\text{i} n\theta_1} , \quad J=1, \label{subeq1}
\\
f_n  &=  - \frac{\kappa}{2\pi a}  \Big( \frac{\tau^2 - 1 -  \text{e}^{-\text{i} n \pi}}{\tau^2 - 2} \Big) \big( w(\theta_1)  +w(\theta_1 + \pi) \text{e}^{-\text{i} n \pi}\big)  \text{e}^{-\text{i} n\theta_1} ,
\notag 
\\
&= - \frac{\kappa}{2\pi a} \sum\limits_{j=1}^{2} w(\theta_j) \text{e}^{-\text{i} n\theta_j}  \times
\begin{cases}
\frac{\tau^2}{\tau^2-2},  &  \ \text{for odd}\ n, 
\\
1 ,& \ \text{for even}\ n, 
\end{cases}  \quad J=2, 
\label{subeq2}
\\
f_n  &=  - \frac{\kappa}{2\pi a} \Big( \frac{1}{\tau^2 - \frac{J}{2}} \Big) \sum\limits_{j=1}^{J} \bigg[ \sum\limits_{m=1}^{J} w(\theta_m) \cos(\theta_j-\theta_m) + \Big(\tau^2 - \frac{J}{2}\Big) w(\theta_j) \bigg] \text{e}^{-\text{i} n\theta_j} ,
\notag 
\\
&= - \frac{\kappa}{2\pi a} \sum\limits_{j=1}^{J} w(\theta_j) \text{e}^{-\text{i} n\theta_j}  \times
\begin{cases}
\frac{\tau^2}{\tau^2-\frac{J}{2}},  & n= \pm 1 \,\text{mod}\, J, 
\\
1, & \text{otherwise} , 
\end{cases} 
\qquad J\ge 3, 
\label{subeq3}
\end{align}
\end{subequations}
where the results \eqref{w_proj} and \eqref{-345} were used for $J\ge 3$ axisymmetrically distributed springs with $\theta_{m+1} = \theta_m + 2\pi/J$ \rev{and the notation $n= \pm 1 \,\text{mod}\, J$ is defined in eq.~\eqref{nmod}.} The modal force for $J$ springs is of only two types. The solution with the coefficient ${\tau^2}/(\tau^2-\frac{J}{2})$ is the same as for the single spring. The dependence on $\tau^2$ (i.e. the mass $m$) implies that the displacement of the internal mass contributes to the net modal force for modes $n= \pm 1 \,\text{mod}\, J$. The second solution is independent of $m$, suggesting that although the mass does displace as seen in \eqref{0-5}, there is no net force on the shell due to this displacement.

Equations \eqref{_25} indicate that the set of force coefficients $\{f_n  \}$ depend upon $J$ linearly independent combinations of the the radial displacements $\{ w(\theta_j)\}$. Thus, for $J=1$ we have $w(\theta_1)$ only; for $J=2$ 
it is $w(\theta_1)+w(\theta_2) $ and $w(\theta_1)-w(\theta_2) $; for $J=3$ 
we have $w(\theta_1)+w(\theta_2)+w(\theta_3) $, $w(\theta_1)\text{e}^{-\text{i} \theta_1}+w(\theta_2)\text{e}^{\text{i} \theta_1}+w(\theta_3) $, and $w(\theta_1)\text{e}^{\text{i}\theta_1}+w(\theta_2)\text{e}^{-\text{i}\theta_1}+w(\theta_3) $; etc. These independent combination of $\{ w(\theta_j)\}$ can also be represented in terms of the infinite series of Fourier coefficients $\{W_m\}$, see eq.\ \eqref{disp}. Assuming the springs are fixed to the shell at \rev{$\theta_j = j2\pi /J$, $j=1,\ldots,J$}, it follows from eqs.\ \eqref{_25} and \eqref{-345} that the force coefficients can be succinctly expressed %in the form 
\beq{-55}
f_n = - \frac{J\kappa}{2\pi a} \, w_n^{(J)} \times
\begin{cases} 
  \frac{\tau^2}{\tau^2 - 1}  
, & J=1, 
\\
\begin{cases} 
\frac{\tau^2}{\tau^2-2},  &  n \ \text{odd}, 
\\
1 ,& n \ \text{even},
\end{cases}    &J=2, 
\\
\begin{cases}
\frac{\tau^2}{\tau^2 - \frac{J}{2}} ,&   n= \pm 1\,\text{mod}\, J,  
\\
1 ,&  \text{otherwise}, 
\end{cases}  &J\ge 3, 
\end{cases}
\eeq
where 
\beq{6-6}
w_n^{(J)} \equiv \sum\limits_{p=-\infty}^{\infty} W_{n+Jp}. 
\eeq
Note that $w_n^{(J)} = w_m^{(J)}$ if $m=n\,$mod$\, J$. Also, note that for a diametrical pair of springs, $J=2$, the summation in \eqref{_25} contains the term $(1+\text{e}^{\text{i}(m-n)\pi})$, which is zero unless $n$ and $m$ are both even or both odd, resulting in $1+\text{e}^{\text{i}(m-n)\pi}=2$. The representation \eqref{-55} for $f_n$ will prove to be crucial for relating the internal dynamics with the external scattering.

\section{Scattering from a shell with an internal substructure}\label{sec4}

\subsection{The forcing coefficients}

Now that we have an expression for the modal force on the shell in terms of the modal displacement we can substitute it into the equation of motion \eqref{eoma} with the replacement $\sigma =  f-p$,  and hence $\sigma_n \to f_n - P_n$. The definition of the shell impedance \eqref{-5} gives $f_n - P_n  = -\text{i} \omega Z_n^{sh} W_n $. Combined with the continuity equation in the form \eqref{2-2}, this yields (see \eqref{8b} for $\zeta_n$)
\beq{7}
W_n =   \frac{2 A_n}{\pi \omega ka \zeta_n} - \frac{f_n} {\text{i} \omega (Z_n^{sh} +Z_n)}.
\eeq
The scattered field is again given by eqs.\ \eqref{Bn} which involves the displacement coefficients $W_n$. It remains to find $W_n$  as a function of the incident wave amplitudes $A_n$.

As shown in the previous section, there are $J$ distinct forms of the modal force $f_n$, each dependent upon the $J-$cyclic parameters $w_n^{(J)} $ of \eqref{6-6}.  These may be determined by taking appropriate summations of \eqref{7}. Define the $J-$cyclic parameters 
\beq{3=4}
\ba
\frac 1{Z_{n}^{(J)}} &=   \sum\limits_{p=-\infty}^{\infty} \frac
1{Z_{n+pJ}^{sh}+Z_{n+pJ}} ,
\quad
{p_{n}^{(J)} }&=  \frac{\text{i} 2 Z_{n}^{(J)} }{\pi ka} \sum\limits_{p=-\infty}^{\infty} \frac{A_{n+pJ}}{\zeta_{n+pJ} }, 
\ea
\eeq
then \eqref{7} implies 
\beq{5=7}
 w_n^{(J)}  = \frac{f_n - p_{n}^{(J)} }{ - \text{i} \omega Z_{n}^{(J)}}  .
\eeq
Equations \eqref{-55} and \eqref{5=7} now provide a pair of equations for $w_n^{(J)} $ and $f_n$. We next consider the solutions for $J=1$, $J=2$ and $J\ge 3$ separately.

\subsubsection{A single spring $(J=1)$}

In this case there is only one modal displacement coefficient $w^{(1)} = w_n^{(1)}$  independent of $n$, as are the force coefficients $f_n$: 
\beq{0=5}
f_n = { \text{i} \omega Z^{sp} w^{(1)} }  \ \ \text{with } \ \ 
w^{(1)} = \sum\limits_{n=-\infty}^{\infty} W_n = \frac{(\text{i} \omega)^{-1} p^{(1)} }{Z^{(1)} + Z^{sp}} 
\eeq
where $p^{(1)} = p_n^{(1)}$, $Z^{(1)} = Z_n^{(1)}$ and 
$Z^{sp}$ 
are 
\begin{equation}
p^{(1)} =\frac{\text{i} 2Z^{(1)} }{\pi ka} \sum\limits_{n=-\infty}^{\infty} \frac{A_n}{\zeta_n}, 
\quad
\frac 1{Z^{(1)}}=  \sum\limits_{n=-\infty}^{\infty} \frac
1{Z_n^{sh} +Z_n}    ,
\quad
Z^{sp} = \frac{\text{i} \kappa}{2 \pi a \omega}\,  \frac{\tau^2}{\tau^2 - 1}.
\end{equation}
See eq.\ \eqref{02} for $Z_n^{sh}$ and eq.\ \eqref{3=3} for $Z_n$. The effective spring impedance is denoted by $Z^{sp}$ with a resonant frequency $\omega^2 = \omega_{sp}^2$, see \eqref{3333}.

\subsubsection{Diametrical pair of springs $(J=2)$}
Now consider the internal mass being supported by a diametrical pair of springs. The modal force is given by eq.\ \eqref{-55}. Unlike the single spring scenario, here, due to symmetry of the spring positions, odd and even modes engage the internal mass differently. This gives rise to the two solutions, for even and odd $n$, as
\beq{9=0}
f_n  = { \text{i} \omega Z^{sp}_{\stackrel{e}{o}} w^{(2)}_{\stackrel{e}{o}} } 
\ \  \text{with} \ \
w^{(2)}_{\stackrel{e}{o}} = \sum\limits_{n \, \text{even/odd} } W_n = 
\frac{(\text{i} \omega )^{-1} p_{\stackrel{e}{o}}^{(2)} }{Z_{\stackrel{e}{o}}^{(2)} + Z^{sp}_{\stackrel{e}{o}}} 
\eeq
where
\begin{equation} \label{eo}
p_{\stackrel{e}{o}}^{(2)} = \frac{\text{i} 2Z_{\stackrel{e}{o}}^{(2)} }{\pi ka} \sum\limits_{n \, \text{even/odd}}  \frac{A_n }{\zeta_n}, 
\quad
\frac{1}{Z_{\stackrel{e}{o}}^{(2)}}=  \sum\limits_{n \, \text{even/odd}} \frac{1}{Z_n^{sh} +Z_n}    ,
\quad
Z^{sp}_{\stackrel{e}{o}} = \frac{\text{i} \kappa}{\pi a \omega} \times
\begin{cases} 
\frac{\tau^2}{\tau^2 - 2},  &  n \ \text{odd}, 
\\
1 ,& n \ \text{even} .
\end{cases}
\end{equation}
Similar expressions were derived by Guo in \cite{Guo92}. Note, for a diametrical pair of springs, the resonant frequency is $\omega^2 = 2 \omega_{sp}^2$, see \eqref{3333}$_1$.

\subsubsection{Axisymmetric distribution of three or more  springs $(J\ge 3)$}
The solution for $J\ge 3$ axisymmetrically distributed springs is essentially the same as for two springs, namely,
\begin{equation}  \label{solJ3}
f_n = { \text{i} \omega Z^{sp}_{n} w^{(J)}_n } 
\ \ \text{where} \ \ 
w^{(J)}_n = \sum\limits_{p=-\infty}^{\infty} W_{n+pJ} = \frac{(\text{i}\omega )^{-1}p_n^{(J)} }{Z_n^{(J)} + Z^{sp}_n} ,
\end{equation}
with
\begin{equation} \label{J}
p_n^{(J)} = \frac{\text{i} 2Z_n^{(J)}}{\pi ka}  \sum\limits_{p=-\infty}^{\infty} \frac{A_{n+pJ}}{\zeta_{n+pJ}}, 
\quad
\frac{1}{Z_n^{(J)}}=   \sum\limits_{p=-\infty}^{\infty} \frac{1}{Z_{n+pJ}^{sh} +Z_{n+pJ}}   ,
\end{equation}
and
\begin{equation} \label{ZspJ}
Z^{sp}_n = \frac{\text{i} {J} \kappa}{2 \pi a \omega} \times
\begin{cases}
\frac{\tau^2}{\tau^2 - \frac{J}{2}} ,&   n= \pm 1\,\text{mod}\, J,  
\\
1 ,&  \text{otherwise}. 
\end{cases}
\end{equation}
The summation in \eqref{solJ3} is $J$-cyclic, $w^{(J)}_n=w^{(J)}_{n\text{mod}\, J}$. Thus, there are $J$ unique solutions that need to be determined $\{w^{(J)}_0, w^{(J)}_1, w^{(J)}_2, \dots, w^{(J)}_{J-1} \}$, where the spring impedance for $w^{(J)}_1$ and $w^{(J)}_{J-1}$ differs from other solutions as seen in eq.\ \eqref{ZspJ}.

\subsection{Scattering solution}
Write the scattering coefficients from eq.\ \eqref{Bn} 
as 
\beq{1=5}
B_n= B_n^{(0)}  + B_n^{(1)}, 
\eeq
where $B_n^{(0)}$ are the values for system with no internal spring-mass system.  Thus, 
using eqs.\ \eqref{8b} and \eqref{7}, 
\beq{1=4}
B_n^{(0)}  = \frac{1}{2}\left( \frac{\zeta_n^*}{\zeta_n} - 1\right) A_n, 
\qquad
 B_n^{(1)}=  \text{i} \rho c \, \frac{ f_n }{\zeta_n} .
\eeq
 Substituting the forcing coefficient of eq.\ \eqref{solJ3} into eq.\ \eqref{1=4}$_2$, the contribution of the internal spring-mass system to the scattering coefficient is
\begin{equation}
B_n^{(1)} = \text{i} \rho c \, \frac{ Z^{tot}_n\, p_{n}^{(J)} }{Z^{(J)}_{n} \zeta_n}  ,
\qquad 
\frac{1}{Z^{tot}_n} = \frac{1}{Z^{(J)}_{n}} + \frac{1}{Z^{sp}_{n}} ,
\end{equation}
where $Z^{tot}_n $ is the equivalent total impedance of the shell-spring-mass system. In the following subsections, the scattering coefficients and the T-matrices are determined separately for $J=1$, $J=2$, and $J\ge3$ springs.

\subsubsection{Scattering coefficients, $J=1$}
For a single spring, the scattering coefficient is
\begin{equation} \label{4-3}
B_n^{(1)} = - \frac{2\rho c Z^{tot}}{\pi ka \zeta_n}\ \sum\limits_{m=-\infty}^{\infty} \frac{A_m }{ \zeta_m} .
\end{equation}
Equation  \eqref{4-3} can be rewritten compactly by defining the infinite vector ${\bf b}$ with elements $b_n$ as 
\beq{306}
{\bf B}^{(1)} = {\bf b} {\bf b}^T {\bf A}
\quad \text{with}\ \ 
b_n = \frac{\text{i}}{\zeta_n} \left( \frac{2\rho c Z^{tot}}{\pi ka}\right)^{1/2}  .
\eeq
Hence, referring to eq.\ \eqref{-41} where, after truncating the series at $N$, the vectors $\bf B$ and $\bf A $ are
\begin{equation}
{\bf B}_{(2N+1)\times 1} =
\begin{pmatrix}
B_{-N}  \\
B_{-N+1}  \\
...  \\
B_N
\end{pmatrix} ,
\quad
{\bf A}_{(2N+1)\times 1} =
\begin{pmatrix}
A_{-N}  \\
A_{-N+1}  \\
...  \\
A_N
\end{pmatrix}
\end{equation}
and the T-matrix is 
\begin{equation} \label{32}
{\bf T} = {\bf T}^{(0)} + {\bf b} {\bf b}^T , 
\end{equation}
where ${\bf T}^{(0)}$ is the diagonal matrix with elements $T_n$ on the diagonal, see \eqref{8b}. The additional non-diagonal matrix in \eqref{32}  is caused by the spring-mass system. %See appendix for explicit form.

\subsubsection{Scattering coefficients, $J=2$}
Recall  that for a diametrical pair of springs there are two solutions for even and odd modes, see  \eqref{9=0}.  The scattering coefficient for even and odd modes, respectively, is (see \eqref{eo})
\begin{equation} \label{4-3b}
(B_n^{(1)})_{\stackrel{e}{o}} = - \frac{2\rho c Z^{tot}_{\stackrel{e}{o}}}{\pi ka \zeta_n} 
\sum\limits_{m \, \text{even/odd}} \frac{A_m }{\zeta_m} ,
\quad \quad
\frac{1}{Z^{tot}_{\stackrel{e}{o}}} = \frac{1}{Z^{(J)}_{\stackrel{e}{o}}} + \frac{1}{Z^{sp}_{\stackrel{e}{o}}} .
\end{equation}
In order to express the scattering coefficient vector in the succinct form  
\begin{equation}
{\bf B}^{(1)} = {\bf b}_e {\bf b}_\text{e}^T {\bf A}
+ {\bf b}_o {\bf b}_o^T {\bf A}
\end{equation}
define the (infinite) vectors ${\bf b}_e$ and ${\bf b}_o$ 
\beq{-278}
{\bf b}_{e} =
\begin{pmatrix}
\vdots  \\
b_{-2}  \\
0  \\
b_{0}  \\
0  \\
b_{2}  \\
\vdots  \\
\end{pmatrix},
\quad
{\bf b}_{o} =
\begin{pmatrix}
\vdots   \\
0 \\ 
b_{-1}  \\
0  \\
b_{1}  \\
0  \\
\vdots   \\
\end{pmatrix}
, \quad (b_n)_{\stackrel{e}{o}} = \frac{\text{i}}{\zeta_n} 
\left(  \frac{2\rho c Z^{tot}_{\stackrel{e}{o}}}{\pi ka}\right)^{1/2} .
\eeq
Thus the T-matrix becomes
\begin{equation} \label{32J2}
{\bf T} = {\bf T}^{(0)} + {\bf b}_e {\bf b}_\text{e}^T  
 + {\bf b}_o {\bf b}_o^T  .
\end{equation}
The structure of the T-matrix in \eqref{32J2} is very interesting. It means that the additional scattering above and beyond that of the shell without the spring-mass is of only two types, proportional to ${\bf b}_e$ or ${\bf b}_o$.  The amplitude of each type of scattered field depends on how the incident wave couples to it, and this is given by the inner products ${\bf b}_\text{e}^T {\bf A}$  and ${\bf b}_o^T {\bf A}$.

We note that the influence of the spring-mass enters through the two frequency dependent impedances $Z^{sp}_e$ and $Z^{sp}_o$.  They couple to the shell  and the radiating wave impedances, $Z^{sh}_n$ and $Z_n$ in series via the expressions in \eqref{eo}.

\subsubsection{Scattering coefficients, $J\ge 3$}
In the general case of $J\ge 3$ springs, the scattering coefficient is (see \eqref{J} and \eqref{ZspJ})
\begin{equation} \label{4-3c}
B_n^{(1)} = - \frac{2\rho c Z^{tot}_n}{\pi ka \zeta_n}\ 
\sum\limits_{p=-\infty}^{\infty} \frac{A_{n+pJ} }{ \zeta_{n+pJ}} ,
\quad \quad
\frac{1}{Z^{tot}_n} = \frac{1}{Z^{(J)}_n} + \frac{1}{Z^{sp}_n} .
\end{equation}
Conveniently, the vector of scattering coefficients  can be written as
\begin{equation}
{\bf B}^{(1)} = {\bf b}_1 {\bf b}_1^T {\bf A}
+ {\bf b}_2 {\bf b}_2^T {\bf A} + \ldots + {\bf b}_J {\bf b}_J^T {\bf A}  ,
\end{equation}
where
\beq{-35}
{\bf b}_{j} =
\begin{pmatrix}
...  \\
b_{j,j-J}  \\
{\bf 0}_{(J-1)\times 1}  \\
b_{j,j}  \\
{\bf 0}_{(J-1)\times 1}  \\
b_{j,j+J}  \\
{\bf 0}_{(J-1)\times 1}  \\
...  \\
\end{pmatrix}
\quad b_{j,n} = 
\begin{cases}
\frac{\text{i}}{\zeta_n} 
\left(  \frac{2\rho c Z^{tot}_{n}}{\pi ka}\right)^{1/2} , &  n=j\,\text{mod}\, J,
\\
0 ,& \text{otherwise}.
\end{cases}
\eeq
The full T-matrix then takes the form
\begin{equation} \label{TJ}
{\bf T} = {\bf T}^{(0)} +  \sum\limits_{j=1}^{J} {\bf b}_j {\bf b}_j^T  . 
\end{equation}

\section{Discussion of the general solution}\label{sec5}
The structure of the derived results is well suited for numerical implementation. The contribution of the spring-mass system to the to the T-matrix of the empty shell is expressed via vectors, thereby removing the need for matrix multiplication. Also, the $J$ sub-solutions only need to be added to produce the final response.

\subsection{Spectral properties of $\bf T$}
Let $\lambda$ be an eigenvalue of the T-matrix with associated  eigenvector $\bf u$, i.e. 
\beq{802}
\bf T   u = \lambda   u . 
\eeq
We note that the equation for $\lambda$, det$\, ({\bf T}   -\lambda {\bf I}) = 0$, can be expressed 
\beq{806}
 {\bf b}_j^T
\big(  \lambda {\bf I} - {\bf T}^{(0)} \big)^{-1}{\bf b}_j  
=  1 , \quad j=1, \ldots, J  .
\eeq
In order to see this, first use \eqref{TJ} to rewrite \eqref{802} as 
\beq{804}
{\bf u} =  \sum\limits_{j=1}^{J} ( {\bf b}_j^T  {\bf u})\,  
\big(  \lambda {\bf I} - {\bf T}^{(0)} \big)^{-1}{\bf b}_j .
\eeq
Taking the inner product with ${\bf b}_i$ yields 
\beq{805}
 \sum\limits_{j=1}^{J}  {\bf b}_i^T
\big(  \lambda {\bf I} - {\bf T}^{(0)} \big)^{-1}{\bf b}_j
\,  ( {\bf b}_j^T  {\bf u}) 
=  {\bf b}_i^T{\bf u}    .
\eeq
This simplifies by virtue of the facts that  $\lambda {\bf I} - {\bf T}^{(0)} $ is diagonal, and that, for any diagonal matrix ${\bf D}$, $ {\bf b}_i^T {\bf D}{\bf b}_j = \delta_{ij}\, {\bf b}_j^T {\bf D}{\bf b}_j$, \rev{where $\delta_{ij}$ is the Kronecker delta and} from which \eqref{806} follows. 

Equation \eqref{806} implies that the eigenvalues of the T-matrix form $J$ distinct sets, and that the eigenvectors, which follow form \eqref{804}, are likewise separated into $J$ families. Hence, $\bf T $ can be partitioned into $J$ distinct T-matrices: 
\beq{807}
\ba
{\bf T} &=  \sum\limits_{j=1}^{J} {\bf T}^{(j)}
\ \ \text{where} \  \ 
 {\bf T}^{(j)} = {\bf T}^{(0)}  {\bf I}^{(j)} +  {\bf b}_j {\bf b}_j^T  ,
\\
{\bf I} &=  \sum\limits_{j=1}^{J} {\bf I}^{(j)},  \ \ 
{\bf I}^{(j)} = \text{diag} \big( 
\ldots 1,\, {\bf 0}_{(J-1)\times 1}, \, 1,\,  {\bf 0}_{(J-1)\times 1} \ldots \big)
%\begin{pmatrix}
%\ddots \\
    %& 1 & & & \text{\large 0} & & \\
    %& & {\bf 0}_{(J-1)\times 1} & & &  \\
    %& & & 1 & & & \\
    %& & & & {\bf 0}_{(J-1)\times 1} & & \\
    %& & \text{\large 0} & & & 1 & \\
    %& & & & & & \ddots
%\end{pmatrix} 
. 
\ea
\eeq
Conservation of energy is ensured in each subset of modes according to 
\beq{70}
 {{\bf S}^{(j)}}^+ {\bf S}^{(j)} =  {\bf S}^{(j)}  {{\bf S}^{(j)}}^+ ={\bf I}^{(j)}
\ \ \text{where} \  \  {\bf S}^{(j)} = {\bf I}^{(j)} + 2 {\bf T}^{(j)} , \ j=1, \ldots, J. 
\eeq

%\beq{808}
%\begin{pmatrix}
%\ddots \\
    %& T_{j-J} & & & \text{\huge0} & & \\
    %& & {\bf 0}_{(J-1)\times 1} & & &  \\
    %& & & T_{j} & & & \\
    %& & & & {\bf 0}_{(J-1)\times 1} & & \\
    %& & \text{\huge0} & & & T_{j+J} & \\
    %& & & & & & \ddots
%\end{pmatrix}
%\eeq

\begin{figure}[H]    %   plots generated by    matlab/plot_J_matrix.m
\centering
\subfloat[\it J\rm=2]{   \includegraphics[width=2.4in] {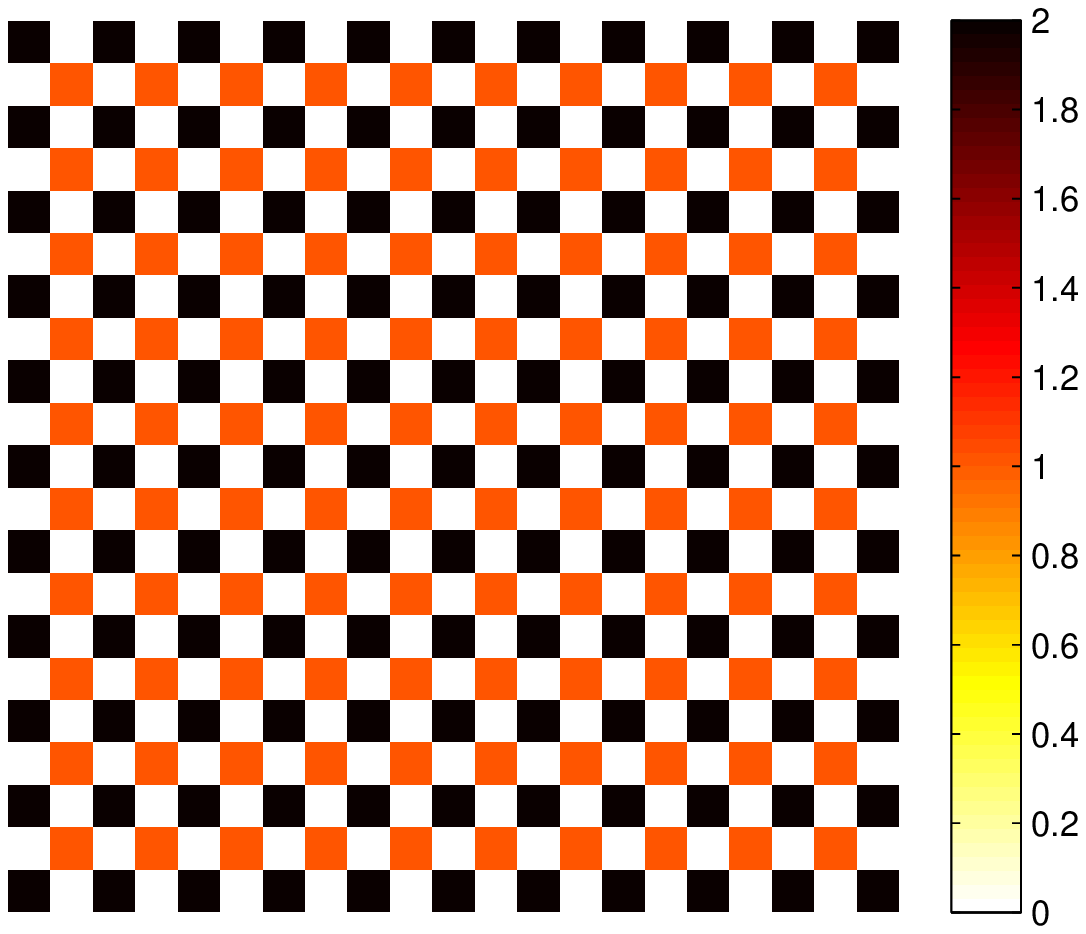}     \label{fig3a}  }
\subfloat[\it J\rm=3]{   \includegraphics[width=2.4in] {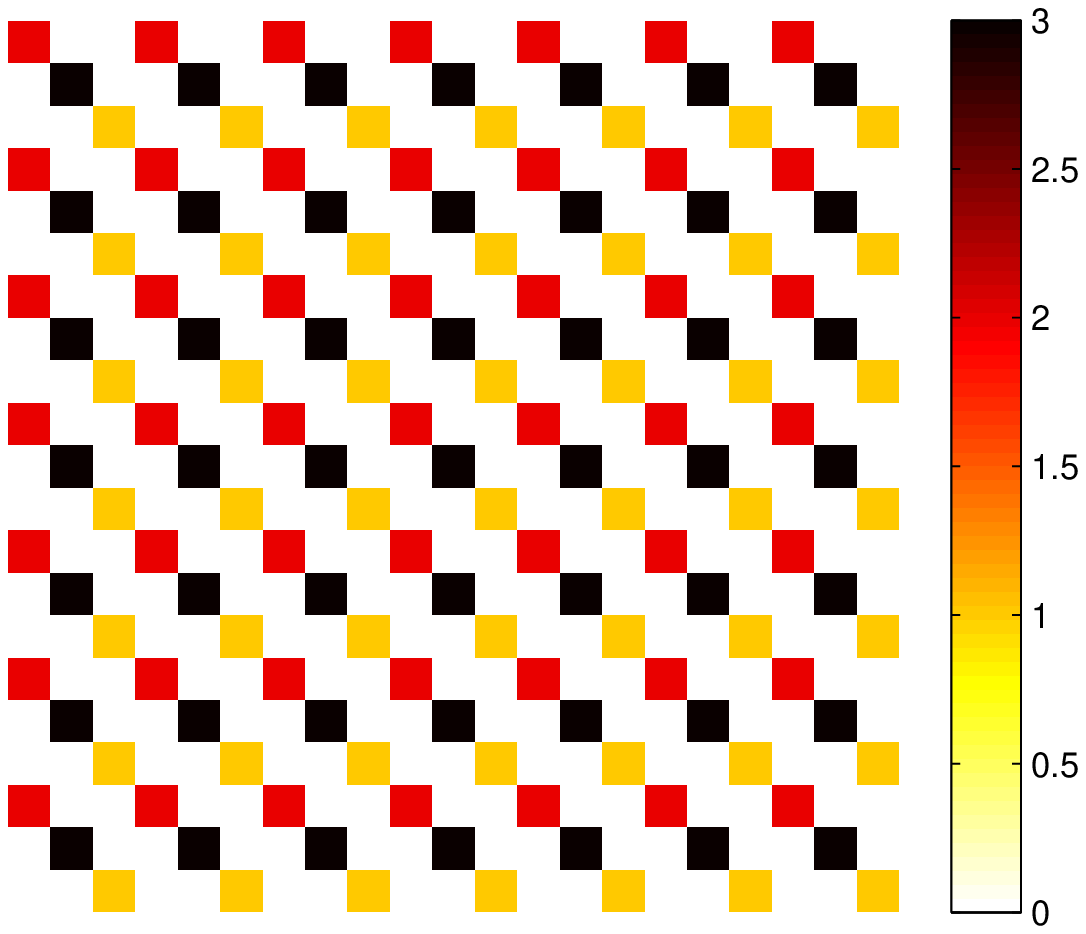}     \label{fig3b}  }
\\
\subfloat[\it J\rm=4]{   \includegraphics[width=2.4in] {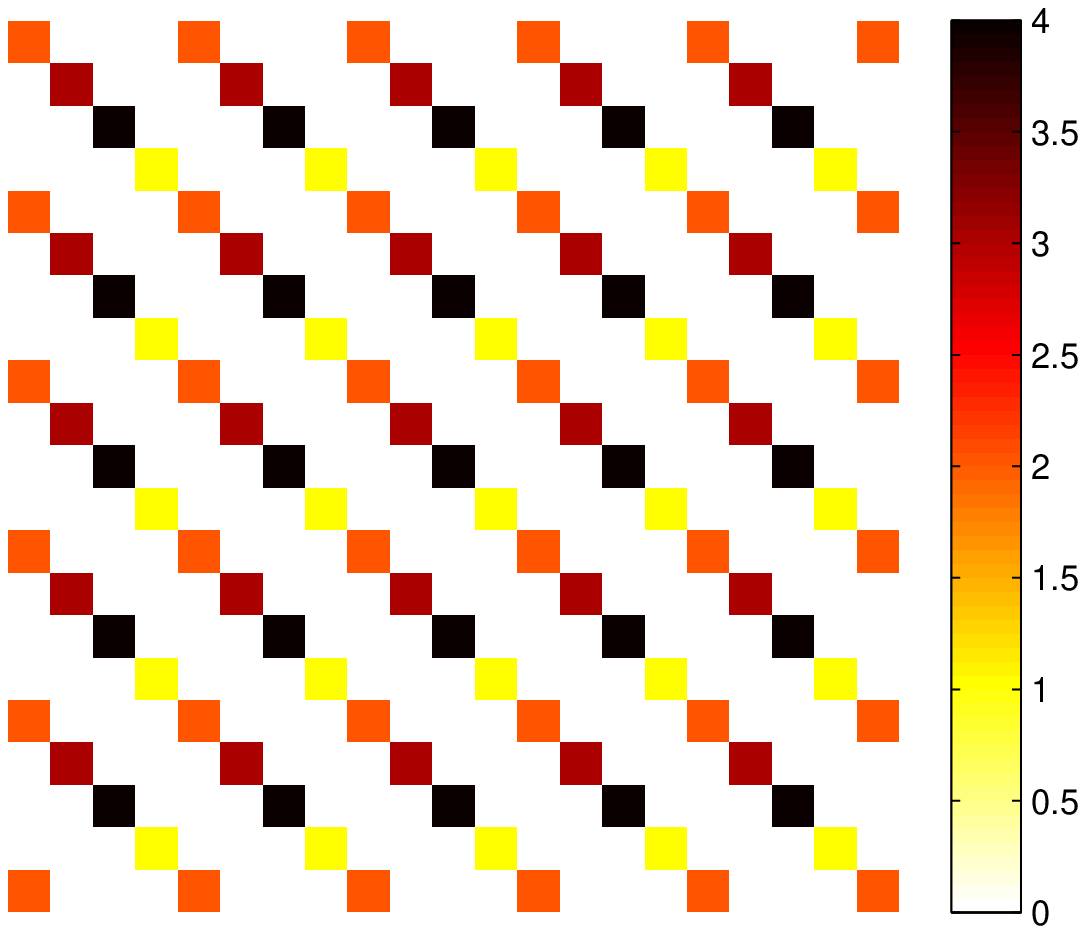}     \label{fig3c}  }
\subfloat[\it J\rm=8]{   \includegraphics[width=2.4in] {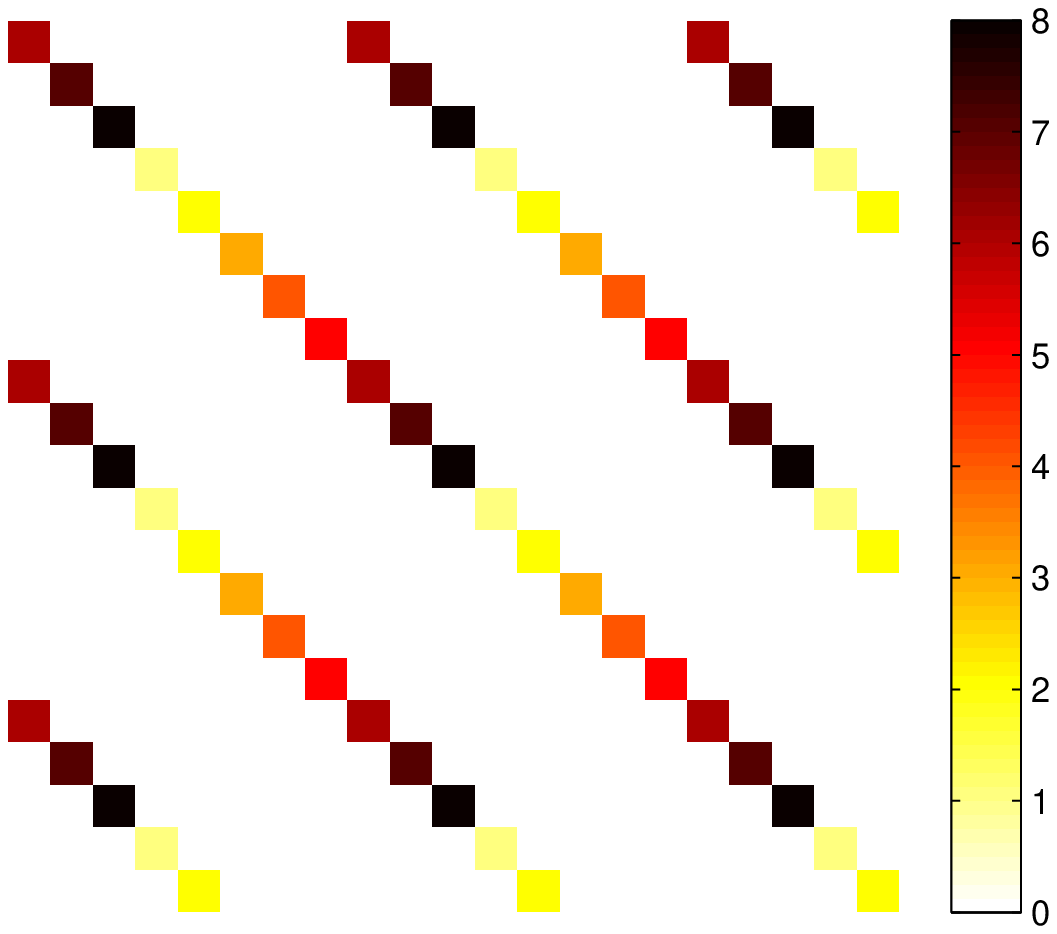}     \label{fig3d}  }
\caption{Structure of the matrices $ {\bf S}^{(j)}$. Matrix elements indicated by  white spaces are zero, and the other colors indicate non-zero elements of the $J$ matrices $ {\bf S}^{(j)}$ for $j=1,\ldots, J$. }
\label{fig3}  
\end{figure}
The structure of these matrices is illustrated in Fig.\ \ref{fig3}. For instance, when $J=2$, Fig.\ \ref{fig3a} shows that one half of the elements of the infinite matrix are zero.  The matrix is full for the case $J=1$, and the number of zero elements increases as $J$ becomes larger. The examples in Fig.\ \ref{fig3} show schematically how the fraction of non-zero elements decreases as $J$ increases: there are always elements on the main diagonal, with the other non-zero elements becoming further separated from the main diagonal as $J$ increases.

\subsection{Far-field Response}
%To obtain the far-field response we employ the large argument approximation of the Hankel function
%\begin{equation}
%H_n^{(1)}(k r)= \sqrt{\frac{2 }{\pi kr}} \text{e}^{-i( n\frac{\pi}{2}+\frac{\pi}{4} )} \text{e}^{i kr} + O\Big( \(kr)^{-3/2}\Big) .
%\end{equation}
The far-field scattered pressure field is 
\beq{-76}
p_s %= \sum\limits_{n = - \infty}^\infty  B_n H_n^{(1)}(kr) \text{e}^{\text{i} n \theta}
 = \sqrt{\frac{a}{2r}} \text{e}^{\text{i} kr} g(\theta) + 
O\Big( ( {kr})^{-3/2}\Big), \quad kr\gg 1,\eeq
where the form function $g$ follows from equation~\eqref{13} and the large argument approximation for Hankel functions,  
\beq{-25} 
g(\theta) =
g^{(0)}(\theta) + g^{(1)}(\theta) = 
\sum\limits_{n = - \infty}^\infty 
g_n  \text{e}^{\text{i} n\theta} ,
\quad 
g_n=  \frac {2\text{e}^{-\text{i} \frac{\pi}{4} } }{\sqrt{\pi ka}}  
(-\text{i})^n   B_n. 
\eeq
The  $g^{(0)}$ is for the shell without the spring system. For a plane wave incident on the shell at an angle $\theta_0$, the scattering coefficient is 
$B_n= \sum_{m = - \infty}^\infty T_{nm}A_m$ 
where $A_m=(-\text{i})^m \text{e}^{-\text{i} m \theta_0}$. This allows us to write the far-field form function as (see \eqref{8b} for $T_n$)
\beq{4-0}
g_n = \frac {2\text{e}^{-\text{i}\frac{\pi}{4} } }{\sqrt{\pi ka}} 
 \sum\limits_{m = - \infty}^\infty 
(-\text{i})^{n+m} \text{e}^{-\text{i}m \theta_0} 
\times \begin{cases}
\big( \delta_{nm}T_m 
+  b_n b_m  \big),  & J=1, 
\\
\big( \delta_{nm}T_m 
+  b^{(e)}_n b^{(e)}_m 
+   b^{(o)}_n b^{(o)}_m \big),  & J=2, 
\\
\big( \delta_{nm}T_m 
+   \sum\limits_{j=1}^{J} { b_{j,n}b_{j,m} } \big) ,  & J {\ge}3,
\end{cases}
\eeq
\rev{where $\delta_{nm}$ is the Kronecker delta}.

A measure of the net radiated power from the shell is the total scattering cross-section (TSCS) $\sigma_{tot} $, and the TSCS for the empty shell  $\sigma_{tot}^{(0)}$, defined as
\beq{sigma_tot}
\sigma_{tot} = \frac{1}{2} \int\limits_{0}^{2\pi} | g(\theta) |^2 d\theta =  \frac{4}{ka} \sum\limits_{n=-\infty}^{\infty} | B_n |^2  ,
\quad \text{and}\ \ 
\sigma_{tot}^{(0)}=\frac{4}{ka} \sum\limits_{n=-\infty}^{\infty} | B_n^{(0)} |^2   .
\eeq

\section{Numerical examples}\label{sec6}

Consider a steel shell ($\rho_s=7810$ kg/m$^3$, $c_p=5505$ m/s) immersed in water ($\rho=1000$ kg/m$^3$, $c=1484$ m/s). Shell thickness to radius ratio is $\frac{h}{a}=\frac{1}{100}$. We define the internal mass to shell mass ratio as $\frac{m}{2 \pi \rho_s ha} = 3$. The spring stiffness is assumed to be such that the resonant frequency of the oscillator satisfies (see \eqref{7=8} for $H_J$)
\begin{equation} \label{kappa}
\frac{\kappa}{m} \frac{a^2}{c^2} = \frac{1}{H_J}
\quad \implies \quad
\kappa = \frac{m}{a^2}  \frac{c^2}{H_J} ,
\end{equation}
which gives $(ka)_{sp}\equiv \sqrt{H_J}\frac{\omega_{sp}a}{c}=1 \; \forall J$.   

\begin{figure} [h!]
\begin{center}
\includegraphics[width=6.5in]{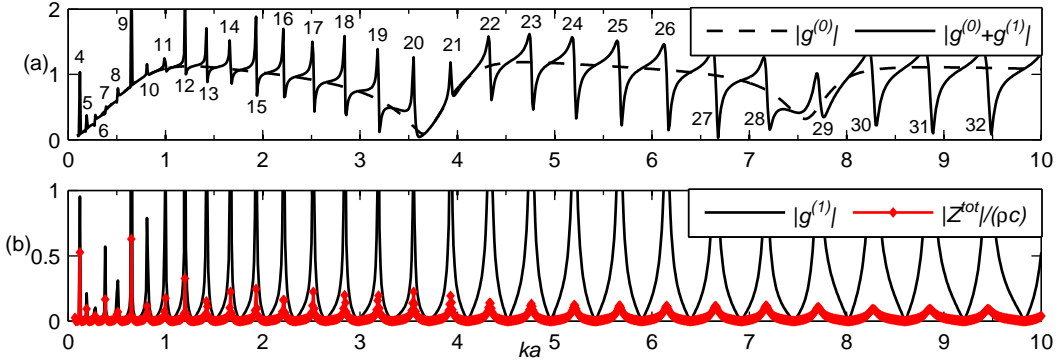}
\caption{Backscatter and total impedance as a function of $ka$ for $J=1$ and incident angle $\theta_0=0$. The dashed line in plot (a) is the backscatter for the empty shell.  In plot (b), the backscatter due to the presence of the spring-mass system $g^{(1)}$ has the same resonances as the total impedance $Z^{tot}$. The small numbers over the resonances indicate the flexural mode.} \label{fig_res_J1}
\end{center}
\end{figure}

\begin{figure} [h!]
\begin{center}
\includegraphics[width=6.5in]{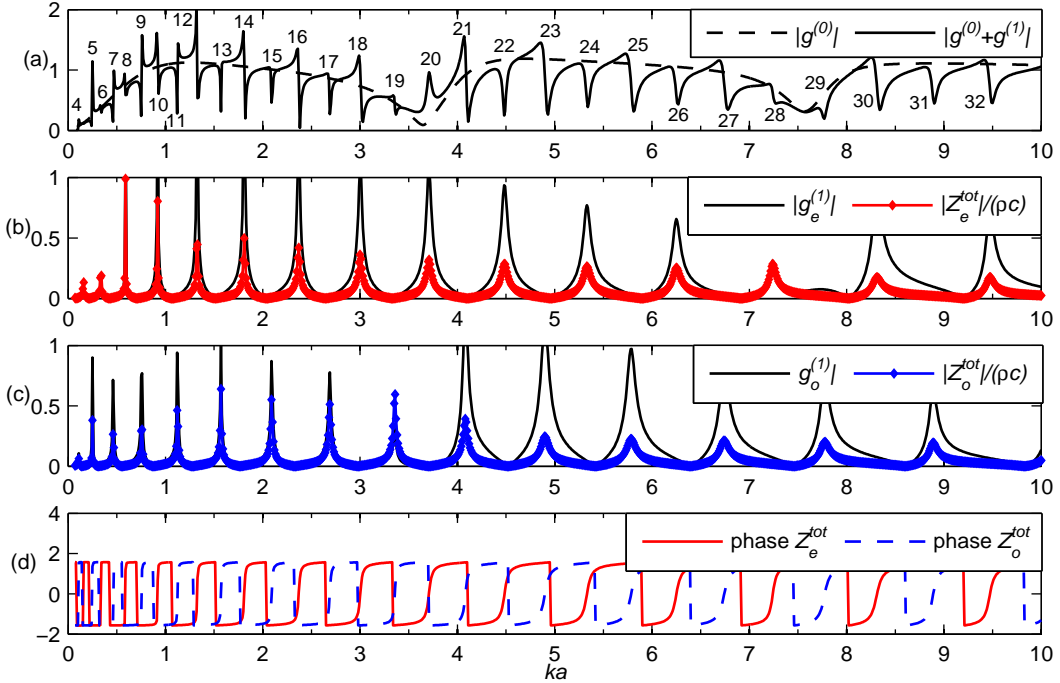}
\caption{Backscatter, total impedance and its phase as a function of $ka$ for $J=2$, $\theta_0=0$. The dashed line in plot (a) is the backscatter for the empty shell. The backscatter due to the even and odd solutions are plotted separately in plots (b) and (c), respectively. The phase of the total impedance is shown in figure (d). The small numbers over the resonances indicate the flexural mode.} \label{fig_res_J2}
\end{center}
\end{figure}

\begin{figure} [h!]
\begin{center}
\includegraphics[width=6.5in]{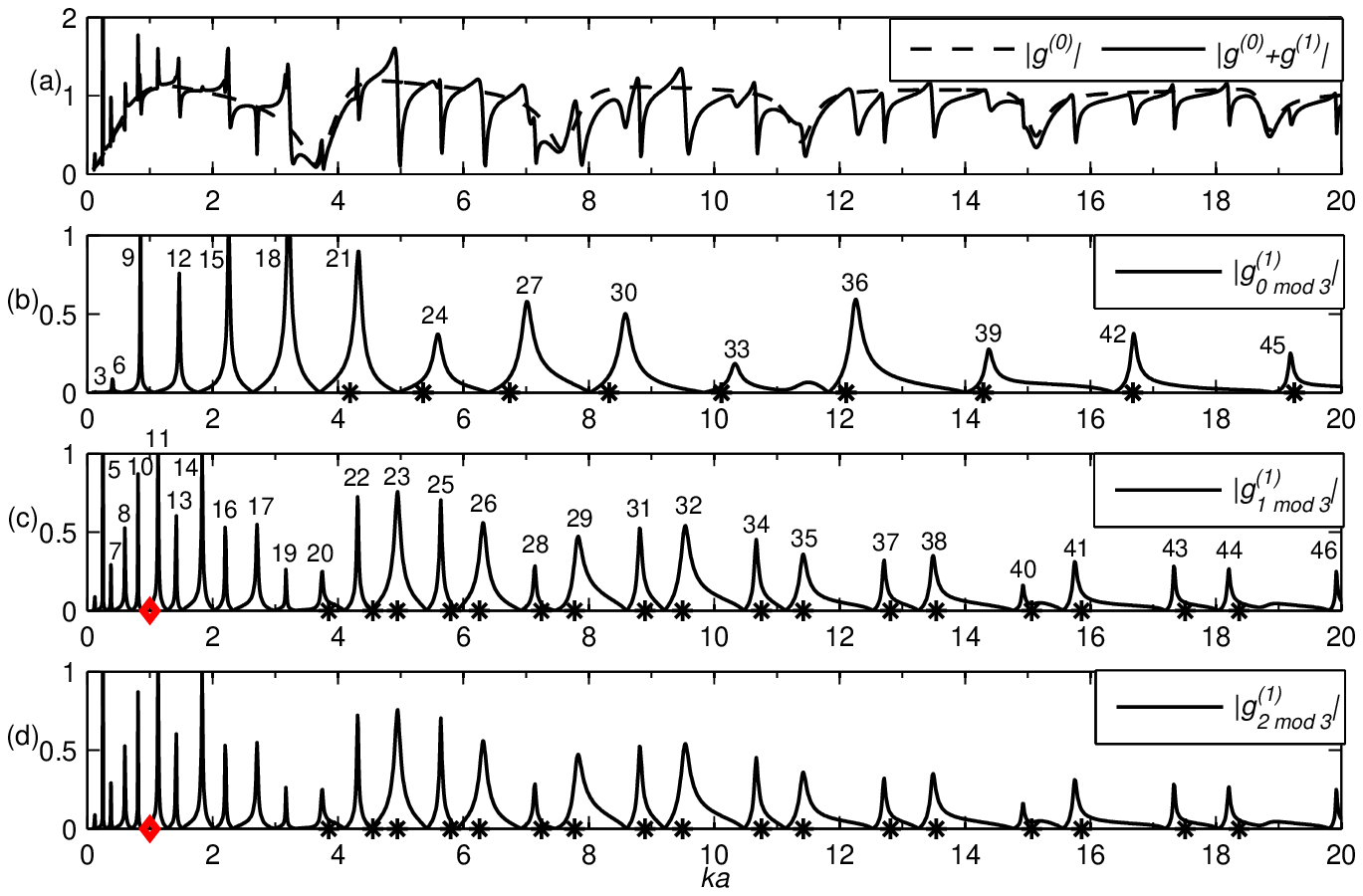}
\caption{Backscatter as a function of $ka$ for $J=3$, $\theta_0=0$. The dashed line in the plot (a) is the backscatter for the empty shell. Plots (b), (c), and (d) show the backscatter for the three subsolutions. The red diamond on the horizontal axis indicates the resonant frequency of the spring-mass system $(ka)_{sp} = 1$. The black stars are the resonances of the shell-spring-mass system as predicted by eq. \eqref{ka_eff}. The small numbers over the resonances indicate the flexural mode. } \label{fig_res_J3}
\end{center}
\end{figure}

Figures \ref{fig_res_J1}, \ref{fig_res_J2}, and \ref{fig_res_J3} show the backscatter $g(\theta_0)$, total impedance $|Z^{tot}|$ and its phase for $J=1,2,3$, respectively. The angle of incidence is taken to be $\theta_0=0$ and the truncation limit is $N=100$. 

With a single spring attaching the internal mass to the shell (Fig.\ \ref{fig_res_J1}) the backscatter is close to that of the empty shell but with many resonances. We show in \S\ref{4.3} that the resonance peaks are associated with flexural modes on the shell excited by the structural discontinuity caused by the spring attachments. The backscatter becomes more complex as the number of springs increases.  For $J=2$ springs, the sub-solutions of the form function are plotted below the total response in Figure \ref{fig_res_J2}. It is evident that half of the resonant peaks come from the even solution and the other half from the odd. At each resonance, the magnitude of the total impedance $|Z^{tot}_{\stackrel{e}{o}}|$ is at a maximum and its phase is zero. This implies that the position and spacing of the resonances can be determined from the total impedance $Z^{tot}$, which is explored further below in \S\ref{4.3}. \rev{The backscatter from the shell with $J=2$ springs in Fig. \ref{fig_res_J2}(a), which was obtained using the general solution (eqs. \eqref{-25} and \eqref{4-0}), is identical to Fig.~3(a) in Ref.~\cite{Guo92}.}

The case of $J=3$ springs in Fig.\ \ref{fig_res_J3} displays a new feature not previously evident for $J=1$ and $J=2$: viz.\ the  $1\, $mod$\, 3$ solution and the $2\, $mod$\, 3$ solution are identical. The repetition is a consequence of (i) the symmetries of the four impedances of eq.\ \eqref{7=8} under the interchange $n\to -n$, and (ii) the fact that the integer sets $1\, $mod$\, 3$ and  $2\, $mod$\, 3$ are identical under a change of sign, i.e.\ $\{\ldots -5,-2,1,4,\ldots \}$ $\leftrightarrow$ $\{\ldots -4,-1,2,5,\ldots \}$. These properties together ensure that the  impedance $Z_n^{tot}$ is also unchanged under $n\to -n$, and hence cause the repetition seen in Fig.\ \ref{fig_res_J3}. It follows that for any $J\ge 1$ the $J$ parts of the $T-$matrix actually reduce to $1+\floor{\frac J2}$ distinct parts, where $\floor {\cdot}$ is the floor function. 

%From the plots for $J=3$ springs (Fig.\ \ref{fig_res_J3}), we observe that the three solutions comprising the total response repeat for $n= 1\,\text{mod}\, J$ and $n= 2\,\text{mod}\, J$. It can be shown that the sub-solutions repeat within the $J$ subsets. As a result, for even $J$ there are only $(J+2)/2$ unique solutions and for odd $J$, $(J+1)/2$. For the current example of $J=3$ springs there are two unique solutions.

\subsection{Resonant behavior of the shell-spring-mass system}\label{4.3}

As noted above for the cases with $J=1$, $J=2$ and $J=3$ springs, the resonant behavior of shell-spring-mass system in Figs.\ \ref{fig_res_J1}-\ref{fig_res_J3} arise from singularities of $Z^{tot}$ lying close to the real $ka$-axis. Thus, at resonance, from eq.\ \eqref{-46}, 
\beq{-34}
\frac{1}{Z^{sp}_{n}}+ 
\sum\limits_{p=-\infty}^{\infty} \frac{1}{Z_{n+pJ}^{sh} + Z_{n+pJ}}   = \epsilon, \ \ |\epsilon| \ll 1.
\eeq
We consider the spring-mass systems of the above numerical examples, for which the resonances are in the range $ka\gg 1$, and in particular, above the spring resonance frequency. The spring impedance is then (see \ref{ZspJ})
\beq{-551}
Z^{sp}_{n} \approx   \frac{\text{i}}{(ka)}\frac{J \kappa}{2 \pi c}  ,  \quad ka\gg 1. 
\eeq
This is independent of $n$, and it's inverse is large, O$(ka)$. We therefore assume that the condition  
\eqref{-34} is satisfied by one of the terms in the infinite series becoming large relative to all others, in which case the condition reduces to 
\beq{res}
Z^{sp}_{n} + Z_{n}^{sh} + Z_{n}  \approx 0 ,  
\eeq
for some $n$ and a related frequency $ka$. The resonances of the combined system are determined by approximating the individual impedances in \eqref{res}. To get an expression for the effective resonant frequencies of the combined system, the acoustic impedance is approximated as
\beq{-57}
Z_n \approx - \text{i} \rho c \frac{ka}{n}, \quad n \gg ka , \quad n \ne 0.
\eeq
The roots of the shell impedance $Z_{n}^{sh}$ of \eqref{02}, which  correspond to the natural frequencies of the shell, are $\Omega_{{c},{f}}^2=((1+n^2+\beta^2 n^4) \pm \sqrt{(1+n^2+\beta^2 n^4)^2-4\beta^2 n^6})/2$ associated with  compressional and flexural modes, respectively. The thin shell approximation implies $\beta \ll 1$, consequently the resonant frequencies are $(\Omega_{c},\Omega_{f})\approx (\sqrt{n^2+1},{\beta n^3}/{\sqrt{n^2+1}})$ while $\beta n < 1$. Since $\Omega_{c}\approx n$, the shell impedance behaves as 
\beq{-56}
Z_n^{sh} \approx -\text{i}\rho_sc_p \frac{h}{a} \left( \Omega-\frac{\Omega_{f}^2}{\Omega} \right) 
\ \  \text{where }\ \ \Omega_f \approx \beta n^2 
\eeq
is the flexural natural frequency.    

%In the intermediate frequency region, the presence of the internal spring-mass system shifts the resonance peaks to the left. 
The condition of resonance given by \eqref{res}, combined with eqs.\ \eqref{-551}, \eqref{-57} and \eqref{-56},  now results in a quadratic equation for the resonant frequencies $(ka)_{res}
= \frac{c_p}c \Omega_{res}$. Solving the equation yields 
\begin{equation} \label{ka_eff}
%(ka)_{res}^2 = \frac{\Omega_{f}^2 + J \hat{\kappa}}{\hat{c}^2 (1 + \hat{\rho} / n )}  
\Omega_{res}^2 = \frac{ \beta^2 n^4   +\frac{J \kappa}{2\pi \rho_s c_p^2} \frac ah }
{1 + \frac 1n \frac{\rho  }{\rho_s  } \frac ah  }
, \quad \quad (ka)_{sp} < (ka)_{res} \ll n .
\end{equation}
%where the wave speed, density, and spring-to-shell stiffness ratios are $\hat{c}=\frac{c}{c_p}$, $\hat{\rho} = \frac{\rho}{\rho_s (h/a)}$, and $\hat{\kappa} =\frac{\kappa}{2\pi \rho_s c_p^2(h/a)}$, respectively.
The resonance can therefore be interpreted as, to leading order, the flexural resonance at $\Omega = \Omega_f $ (see \eqref{-56}) modified by an added mass term in the denominator which accounts for the fluid mass-loading (the same factor is present in eq.\ 9.4 on page 282 of the text by Junger \& Feit \cite{Junger86} for a fluid-loaded spherical shell), and by an additional stiffness term in the numerator associated with the stiffness of the springs.  Note that the flexural resonances are {\it not} excited by the smooth shell (see Fig.\ \ref{fig_res_J1}) because they are sub-sonic and hence do not couple with the incident field. The coupling to the quasi-flexural waves occurs directly because of the introduction of structural discontinuities at the spring-shell attachment points. These act as sources for the flexural waves which, in turn, radiate to the exterior fluid via the same discontinuities.  

\begin{figure} [h!]
\begin{center}
\includegraphics[width=6.5in, height=3.0in]{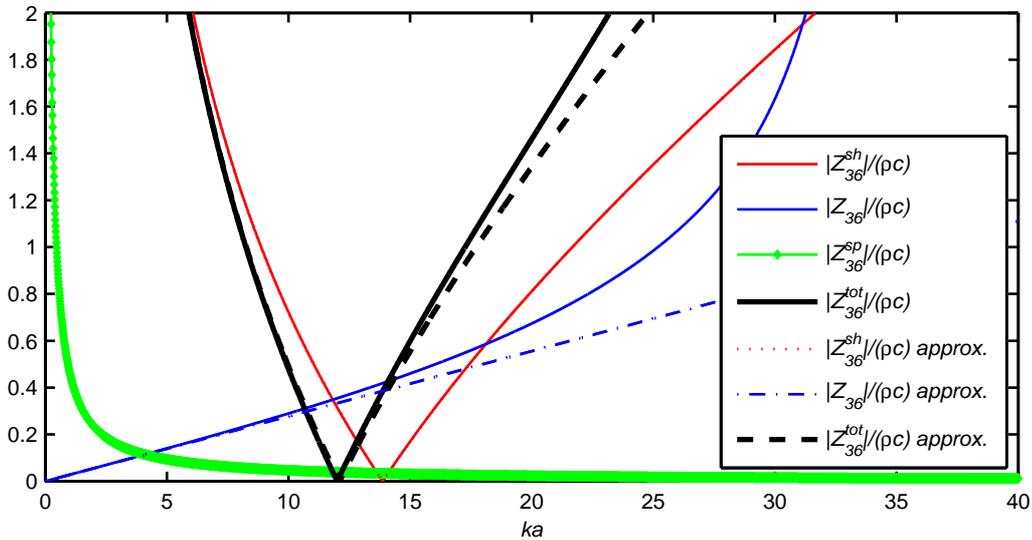}
\caption{Comparison of impedances for mode n=36. This corresponds to the resonance near $ka=12$ of Figure \ref{fig_res_J3}. The solid lines are the exact impedances while the dashed lines represent our approximations.} \label{imped}
\end{center}
\end{figure}

The various approximations leading to the expression for the resonance frequency are verified  in Fig.\ \ref{imped} which shows the approximate impedances plotted along with the exact impedances. The curves are very close as long as $n \gg ka$. At larger frequencies this condition is violated and the expression for the effective frequencies, eq.\ \eqref{ka_eff} is no longer  accurate. However, the spring impedance $Z^{sp}_{n} \sim O(1/(ka))$ and hence its effect at larger frequencies is negligible. The effective resonances are plotted on the horizontal axis for the $J=3$ case in Fig.\ \ref{fig_res_J3}. Although, the values are close to the resonances of the combined system they are not exact. This is primarily because we only take a single term from the summation of $Z^{tot}_n$ when formulating the condition of resonance (see eq. \eqref{-34} and eq. \eqref{res}).

\begin{figure} [h!]
\begin{center}
\includegraphics[width=6.5in]{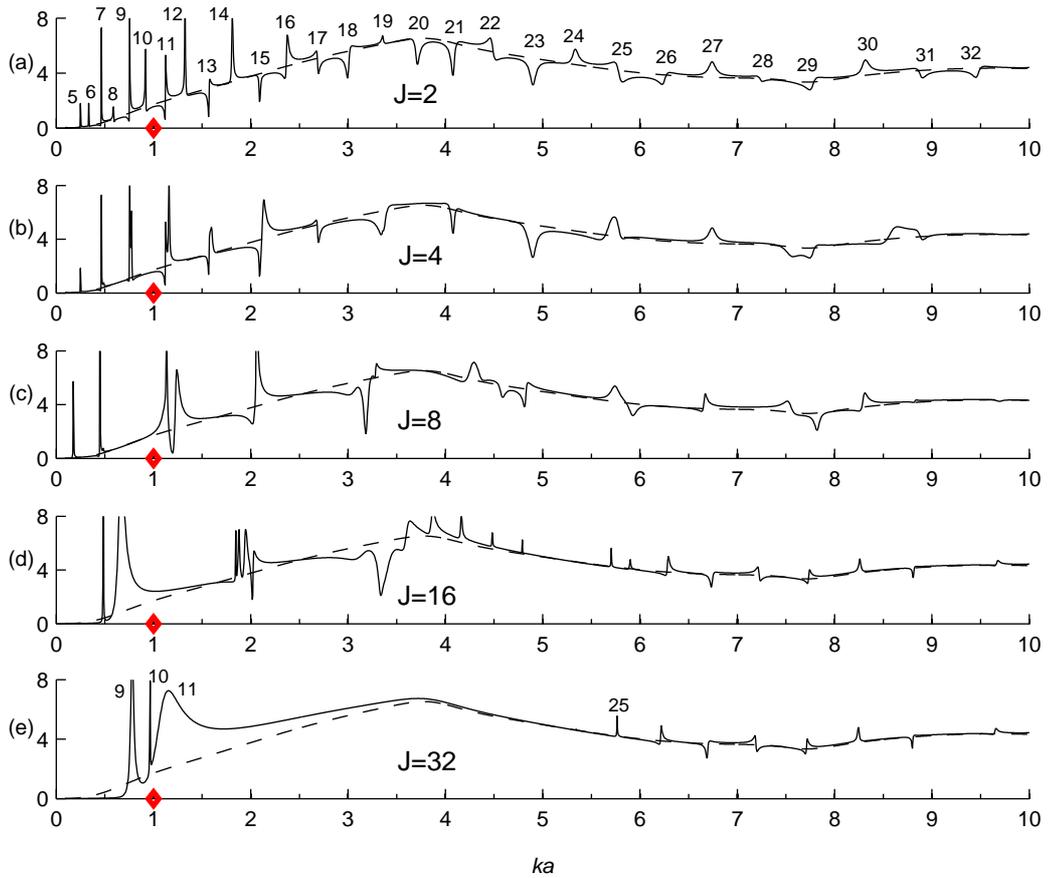}
\caption{\rev{Total scattering cross section as defined in eq.\ \eqref{sigma_tot} for $J=2,4,8,16,32$ springs at the angle of incidence $\theta_0=0$ in plots (a),(b),(c),(d),(e), respectively. The dashed line in all plots is the TSCS for the empty shell. The red diamonds on the horizontal axis indicate the constant resonant frequency of the spring-mass system $(ka)_{sp} = 1$. The small numbers over the resonances indicate the flexural mode.}} \label{larger_J}
\end{center}
\end{figure}

\rev{\subsection{Large $J$ limit}
As the number of springs $J$ increases, the loading on the shell transitions from discrete point forces to an effective pressure at the frequencies of interest $ka\le20$. However, unlike a fluid filled shell where there is also a pressure distribution over the inner surface, in this idealized model the internal structure has an infinite wave velocity since the transfer of energy from one contact point on the shell to the other is instantaneous. Figure~\ref{larger_J} plots the total scattering cross section for the same shell but with increasing number of springs $J=2,4,8,16,32$ at the angle of incidence $\theta_0=0$. Since the resonant frequency of the oscillator is kept constant (see eq.~\eqref{kappa}), the stiffness of each spring has to decrease with increasing $J$. This allows us to investigate solely the affect of increasing the number of contact points.
}

\rev{In general, Fig.~\ref{larger_J} shows that increasing the number of contact points results in a decrease in the number of flexural resonances propagating into the far-field. This is due to the presence of forces at anti-nodes of flexural modes which inhibit their vibration. Only the response for even numbers of springs is plotted and hence the odd modes are prominent as $J$ increases. For $J=16$ and $J=32$, large intervals appear without flexural resonances, however, the TSCS is slightly increased over the empty shell (shown by the dashed line) due to the added stiffness and mass. The low frequency TSCS is asymptotically zero for these two cases because the effective quasi-static properties of the shell-spring-mass system are water-like. For the case with $J=32$ springs there are only a few large resonances near the resonant frequency of the oscillator $k_{sp}a=1$, the $n=9$ and $n=11$ flexural modes. 
}

\rev{As $J$ approaches infinity no flexural modes will be visible in the far-field with the exception of the closest ones to the springs-mass resonance. The mechanically equivalent system as $J\to \infty$ is one of a highly anisotropic medium, with zero azimuthal stiffness and infinite wave speed in the radial direction. The latter is a result of ignoring the spring mass; this could be included but is beyond the goals of the present analysis which is aimed at the low to moderate frequency regime. }

\subsection{Angle of incidence}
\begin{figure}
\centering
\subfloat[\it J\rm=2]{   \includegraphics[width=3.in] {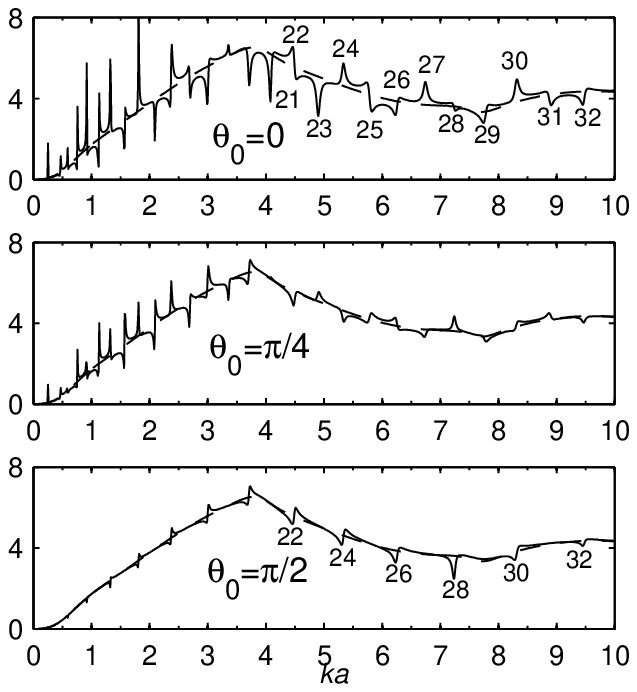}     \label{sigJ2}  }
\subfloat[\it J\rm=3]{   \includegraphics[width=3.in] {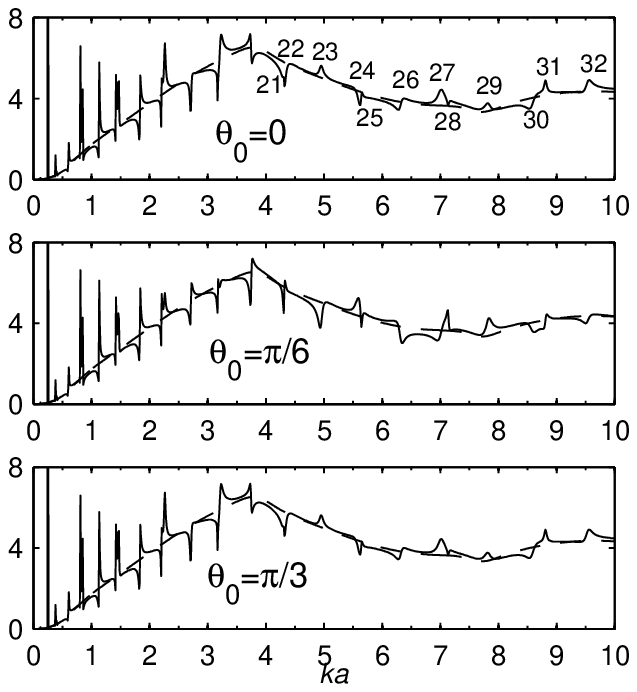}     \label{sigJ3}  }
\\
\subfloat[\it J\rm=4]{   \includegraphics[width=3.in] {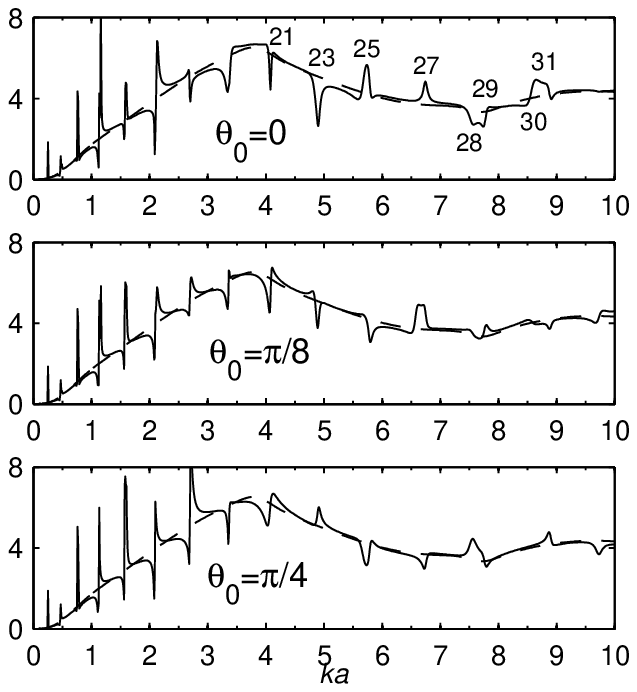}     \label{sigJ4}  }
\subfloat[\it J\rm=8]{   \includegraphics[width=3.in] {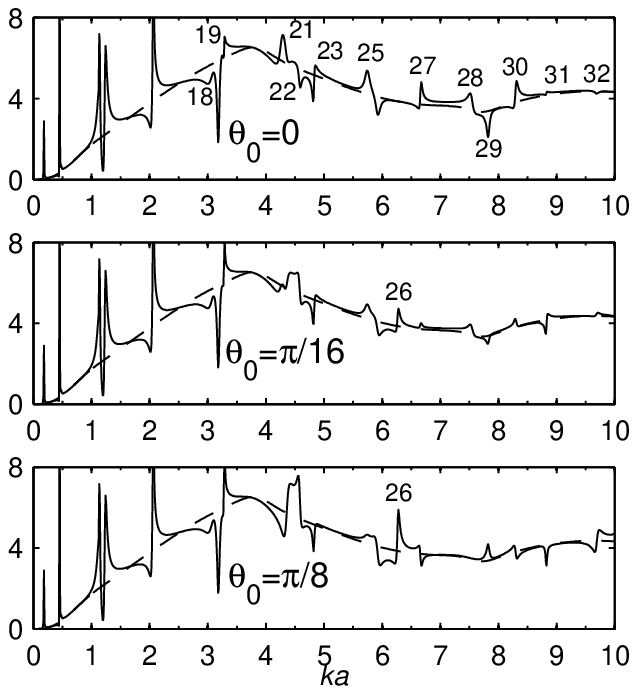}     \label{sigJ8}  }
\\
\subfloat[]{   \includegraphics[width=6.in] {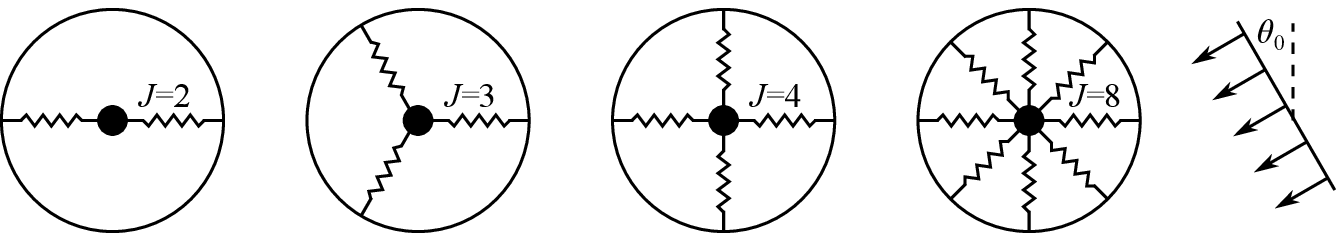}     \label{orientation}  }
\caption{Total scattering cross section as defined in eq.\ \eqref{sigma_tot} for $J=2,3,4,8$ as a function of the plane wave angle of incidence, $\theta_0$. The dashed curve is the total scattering cross section of the empty shell. The spring orientations relative to the incoming wave are shown in figure (e). The spring resonant frequency is $(ka)_{sp}=1$ in all cases.  The small numbers over the resonances indicate the flexural mode.}
\label{fig_sigma}  
\end{figure}

The discrete number of attachment points on the shell produces symmetries which couple to the angle of incidence. The $J$ springs are distributed axisymmetrically, therefore only angles of incidence in the range $\theta_0=[0,\pi / J]$ produce unique results for even $J$. Figure \ref{fig_sigma} presents the total scattering cross section (TSCS) for several distributions of springs. The dashed line represents the TSCS of the empty shell.

For the $J=2$ case in plot (a) we observe a decrease in the amplitude and the number of peaks as the angle of incidence goes from $0$ to $\pi/2$. This is because only the even flexural modes are unconstrained by the springs when the plane wave is perpendicular to the pair as described by \cite{Guo92}. Analysing larger numbers of springs, it has been determined that such clear separation of response also occurs for $J=4$ springs in plot (c). \rev{Furthermore, for $J=2$ springs, as the angle of incidence is changed from $0$ to $\pi/4$, the asymmetric profile of the flexural resonances flips due to a relative phase between the shell-spring-mass system and the surrounding water. In the new results with $J=3$ springs all flexural modes propagate into the far-field regardless of the angle of incidence.} The TSCS for $\theta_0=0$ is identical to that of $\theta_0 =\pi/3$ because both coincide with the orientation of exactly one of the springs.

\rev{The TSCS for $J=4$ springs is shown in plot (c) of Fig.~\ref{fig_sigma}. Again, the total number of peaks is halved as compared to the $J=2$ case due to the fact that the $4$ springs exactly coincide with the anti-nodes of the even flexural modes. Thus the shell stiffened with $J=4$ springs vibrates mostly with odd modes. Because different modes are affected differently by the substructure, certain even modes fall near an excited odd mode and are consequently enhanced. The resulting resonance is no longer sharp, but has a plateau-like form as seen at $ka=7.7$ and $ka=8.7$ for the $n=28,29$ and $n=30,31$ mode pairs, respectively. For $J=8$ springs we see that the low frequency flexural modes are unaffected by the angle of incidence, but the higher modes are affected. For example the $n=26$ flexural mode at $ka=6.28$ is not exited with $\theta_0=0$ but is clearly visible at $\theta_0=\pi/16$ and $\pi/8$.
}

%Comparison with these 
%\subsubsection{Guo's paper}
%Shell thickness is taken as $\frac{a}{100}$, where $a$ is the shell radius. Steel in water. Internal-to-shell mass ratio is 3. The spring stiffness is chosen by defining
%\begin{equation}
%\frac{\kappa}{m} \frac{a^2}{c^2} = 1 .
%\end{equation}
%Thus $\kappa=326\times 10^3 kg/m/s^2$. This puts the resonance in the mid-frequency domain.
%
%\subsubsection{Achenbach's paper}
%Thickness to radius ratio $\frac{h}{a}=\frac{1}{100}$. Mass ratio $\frac{m}{\rho a h \pi}=0.4$. Loss factor defined by $E=E_0(1-1/2i\eta_s)$, with $\eta_s=0.04$. Steel in water. Wave speed ratio defined to be $\frac{c_p}{c}=3.71$ and density ratio defined as $\frac{\rho}{\rho_s} = 7.81$.

\section{Conclusions}   \label{sec7}
The acoustic scattering is determined for a thin elastic shell with an internal mass attached by $J$ axisymmetrically distributed springs. The contribution of the internal system to the T-matrix is comprised of $J$ sub-solutions. Each sub-solution does not represent a single spring, but rather a portion of the combined effect of all springs. The spring attachments on the shell are shown to excite the shell's flexural modes, which are sub-sonic for an empty shell. Using the total impedance, an approximate expression was derived for the resonance frequencies of the fluid-loaded shell-spring-mass system for arbitrary $J$.

The new result, which might not have been clear in the study of systems with $J=1$ and $J=2$ springs, is the relationship between the spring attachment points and nodes/anti-nodes of the flexural modes. We showed that if the spring is attached at the anti-node of a flexural mode, that mode is constrained and the shell is expected to vibrate with modes in which the node is closest to the spring attachment points.  \rev{As $J$ increases the effect of individual springs diminishes. The scattering cross-section becomes asymptotically zero at low frequencies and slightly increases at moderate frequencies due to the added stiffness and mass. The variation of the total scattering cross-section with the angle of incidence is small especially for large $J$.}

\rev{The axisymmetry and tunability of the shell-springs-mass system suggests that it would serve well as a unit cell in fluid-saturated array (sonic crystal) for application in acoustic metamaterials. The low frequency transparency (zero TSCS) at $ka<0.6$ is particularly interesting since this is a steel shell}. The mass and stiffness of each spring were chosen at random, but can be tuned such that the complete system behaves as an effective medium at low frequencies. It is even possible to actively tune such systems by changing the stiffness of the springs. A possible outcome would be an active material for acoustic wave steering, etc. These results will be presented in a separate paper.

\section*{Acknowledgment}
Support from ONR for this work is gratefully acknowledged.

\appendix
\Appendix{3DOF model of a finite sized internal mass} \label{appendixA}
\subsubsection*{One spring} We use Lagrange's equations for the Lagrangian $L=L(x,y,\phi, \dot x, \dot y, \dot \phi) \equiv T-V$, 
%$m\ddot x = - \frac{\partial U}{\partial x}$ etc, 
where $T = \frac 12 m (\dot x^2 + \dot y^2 )+ \frac 12 I\dot \phi^2$ is the kinetic energy, and assuming that the spring is linear, $V= \frac \kappa 2 (l-l_0)^2$, where $l$, $l_0$ are the stretched and un-stretched lengths of the spring. 
\rev{For a spring oriented at angle $\theta_1$ with respect to the positive $x$-axis (refer to Fig.~\ref{fig2}), the   spring length  is given by 
\bal{2}
l^2 =& \ \big|a(\cos\theta_1 , \sin \theta_1) + 
(w\cos\theta_1 -v \sin \theta_1,  w \sin \theta_1 +v \cos \theta_1) 
\notag \\
 & \qquad 
- (x,y) - b(\cos (\theta_1 +\phi), \sin  (\theta_1 +\phi)) \big|^2 .
\eal
The Euler-Lagrange equation for $x$, $\frac{\partial L}{\partial x} - \frac{d }{d t} \frac{\partial L}{\partial \dot x} = 0$, is fully nonlinear and can be cast in the following form
\begin{equation} \label{1} 
 %\frac{\partial L}{\partial x} - \frac{d }{d t}  \frac{\partial L}{\partial \dot x} = 0 \ \  \Rightarrow \ \ 
m\ddot x = -\kappa (l-l_0) \frac{\partial l}{\partial x}  \quad \Rightarrow \quad m\ddot x = -\kappa \frac{(l^2-l^2_0)}{2l(l+l_0)} \frac{\partial l^2}{\partial x} ,
\end{equation}
where $\partial l^2 /\partial x$ follows from \eqref{2}.  
Similar equations for $\ddot y$ and $\ddot \phi$ can be found from the respective  Euler-Lagrange equations. 
%\notag \\
%=& \  \ \ 
%\big(a\cos\theta_1 - b\cos (\theta_1 +\phi) -x +w\cos\theta_1 -v \sin \theta_1\big)^2
%\notag \\  
 %& +
%\big( a\sin \theta_1  -b \sin  (\theta_1 +\phi) -y +  w \sin \theta_1 +v \cos \theta_1 \big)^2 ,
}

\subsubsection*{Linearization} 
\rev{Equations for $\ddot x$, $\ddot y$ and $\ddot \phi$ such as eq.~\eqref{1} form a set of coupled nonlinear ordinary differential equations. In order to get the linear equations we need only the terms  linear in $x,y,\phi$ and $w,v$, or equivalently, linear in $l^2-l^2_0$.} Hence,  
\beq{-1}
 m\ddot x \approx    \frac{-\kappa}{4l_0^2} \left. \frac{\partial l^2}{\partial x}\right|_0  (l^2-l^2_0), 
\ \ 
 m\ddot y \approx    \frac{-\kappa}{4l_0^2} \left. \frac{\partial l^2}{\partial y}\right|_0  (l^2-l^2_0), 
\ \ 
 I\ddot \phi \approx    \frac{-\kappa}{4l_0^2} \left. \frac{\partial l^2}{\partial \phi }\right|_0  (l^2-l^2_0), 
\eeq
where $\left. \right|_0$ indicates the unstretched value $(x=y=\phi=0, w=v=0)$. \rev{Equation~\eqref{2} implies}
\beq{3}
l_0 =a-b,
\ \ 
\left. \frac{\partial l^2}{\partial x} \right|_0 = - 2 l_0 \cos\theta_1 ,
\ \ 
\left. \frac{\partial l^2}{\partial y} \right|_0 =  - 2 l_0 \sin\theta_1 ,
 \ 
\left. \frac{\partial l^2}{\partial \phi}\right|_0 =  0 
\eeq
and
\bal{31}
l^2 - l_0^2 &\approx 
\left. \frac{\partial l^2}{\partial x} \right|_0 x + 
\left. \frac{\partial l^2}{\partial y} \right|_0 y + 
\left. \frac{\partial l^2}{\partial \phi}\right|_0 \phi 
 + 
\left. \frac{\partial l^2}{\partial w} \right|_0 w
 + 
\left. \frac{\partial l^2}{\partial v} \right|_0 v
\notag \\
&= - 2 l_0 (x \cos\theta_1 + y \sin\theta_1) +2l_0 w.
\eal
The linearized equations are therefore 
\beq{4} 
\ba
 m \big(\ddot x ,\, \ddot y\big) &=-  \kappa(x \cos\theta_1 + y \sin\theta_1 -w)
\, \big(\cos\theta_1, \,  \sin\theta_1 \big),
\\ 
I\ddot \phi &= 0 .
\ea
\eeq
The contribution of the rotation angle $\phi$ of the internal mass to the spring force is nonlinear and does not appear in this linear formulation. \rev{As an aside, this will be demostrated by determining the equation of rotational motion of the mass $I\ddot \phi = r\times F$ with all other displacements constrained: $x=0$, $y=0$, $v=0$ and $w=0$. For small displacements $\phi\ll 1$, the vector $r$, which defines the position of the force vector $F=\kappa \frac{|l|-|l_0|}{|l|}l$, is $r=(b,b\phi)$. The deformed spring length vector is $l=(a-b,-b\phi)$ yielding a spring extension of $|l|-|l_0|=(a-b)\sqrt{1+b^2\phi^2/(a-b)^2}-(a-b)$. Using the binomial theorem for the square root, we get $(|l|-|l_0|)/{|l|}\approx\frac 12 {b^2\phi^2}/(a-b)^2$. Lastly, the cross product is $r\times l=-ab\phi$ giving a moment on the mass $r\times F = -\frac 12 {\kappa a}(b\phi)^3 /(a-b)^2$. This demonstrates that due to the geometry of this problem, the contribution of the rotation angle $\phi$ to the spring force is proportional to $\phi^2$ and the contribution to the moment is proportional to $\phi^3$. Thus, in the linearized equations \eqref{4}, we obtain $\phi=0$.}
%Assuming time-harmonic motion and solving for $x,y$ gives
%\beq{87}
%\begin{pmatrix}
%x \\ y 
%\end{pmatrix}= 
%\frac{-w}{\tau^2 - 1} \begin{pmatrix}
%\cos\theta_1  \\ \sin \theta_1  
%\end{pmatrix} 
%, \;\;\;\;\;   .
%\eeq

{
\subsubsection*{$J$ springs} 

The equations of motion in the presence of $J$ springs are
\beq{4_2}
\ba
 m\big(\ddot x ,\, \ddot y\big)&=-  \kappa \sum\limits_{j=1}^{J}(x \cos\theta_j + y \sin\theta_j -w(\theta_j)) \, \big(\cos\theta_j, \,  \sin\theta_j \big), 
 \\ 
 I\ddot \phi &= 0 .
\ea
\eeq
%\beq{4_2}
%\ba
 %m\ddot x &=-  \kappa \sum\limits_{j=1}^{J}(x \cos\theta_j + y \sin\theta_j -w(\theta_j)) \cos\theta_j,
%\\ 
 %m\ddot y &=-  \kappa \sum\limits_{j=1}^{J}(x \cos\theta_j + y \sin\theta_j -w(\theta_j))  \sin\theta_j,
 %\\ 
 %I\ddot \phi &= 0 .
%\ea
%\eeq
Again,  $\phi = 0$. For time harmonic motion $(x\to x\text{e}^{-\text{i}\omega t}, \ldots)$  the equations for $x$ and $y$ follow from \eqref{4_2} as
%\beq{4_21}
%\ba
%\tau^2 x - x\sum\limits_{j=1}^{J} \cos^2\theta_j 
%-  y \sum\limits_{j=1}^{J}  \cos\theta_j \sin\theta_j 
%&= -  \sum\limits_{j=1}^{J}  w(\theta_j) \cos\theta_j ,
%\\ 
 %-  x \sum\limits_{j=1}^{J}  \cos\theta_j \sin\theta_j 
%+\tau^2 y - y\sum\limits_{j=1}^{J} \sin^2\theta_j 
%&= -  \sum\limits_{j=1}^{J} w(\theta_j) \sin\theta_j ,
%\ea
%\eeq
%Using $\cos 2\alpha = 2 \cos^2\alpha -1 = 1-2 \sin^2\alpha$, these relations become
\beq{4_22}
\begin{pmatrix}
2\tau^2 -J-C & -S
\\  %& \\ 
-S  & 2\tau^2 - J+C
\end{pmatrix}
\begin{pmatrix}
x \\  %\\ 
y 
\end{pmatrix}
= - 2
\sum\limits_{j=1}^{J} w(\theta_j)
\begin{pmatrix}
\cos\theta_j 
\\ %\\ 
\sin\theta_j 
\end{pmatrix}, 
\eeq
where (see \eqref{3333}) $\tau^2 = \frac{m \omega^2}{\kappa} $  and 
\beq{433}
C+\text{i}S =  \sum\limits_{j=1}^{J}  \text{e}^{\text{i}2\theta_j} . 
%C = \sum\limits_{j=1}^{J}  \cos 2\theta_j ,
%\quad 
%S = \sum\limits_{j=1}^{J}  \sin 2\theta_j  .
\eeq
Solving for $x$ and $y$, 
\beq{4_24}
\begin{pmatrix}
x \\  %\\ 
y 
\end{pmatrix}
= \frac{- 2}{ (2 \tau^2 - J)^2 -C^2-S^2}
\sum\limits_{j=1}^{J} w(\theta_j)
\begin{pmatrix}
2\tau^2 -J +C & S
\\  %& \\ 
S  & 2\tau^2 - J -C
\end{pmatrix}
\begin{pmatrix}
\cos\theta_j 
\\ %\\ 
\sin\theta_j 
\end{pmatrix} .
\eeq
}

{
For $J>1$ we assume that the angles $\{\theta_j\}$ are uniformly distributed, i.e.\ 
$\theta_{j+1} = \theta_j + 2\pi/J$.  Hence $C+\text{i}S = 0$ for all values of $J$ except $J=1,2$, in which cases 
$C+\text{i}S = J \text{e}^{\text{i} 2\theta_1 }$. 
%\beq{4=90}
%C+\text{i}S = 
%\begin{cases}
%J \text{e}^{\text{i} 2\theta_1 }, & J=1,2,
%\\
%0, & J \ge 3. 
%\end{cases}
%\eeq
Solving \eqref{4_24} for the displacements then yields
\beq{0-5}
\begin{pmatrix}
x \\ y 
\end{pmatrix}= 
\frac{-1}{\tau^2 - H_J} 
\, \sum\limits_{j=1}^{J} w(\theta_j) 
\begin{pmatrix}
\cos\theta_j  \\  \sin\theta_j  
\end{pmatrix} , 
\quad
H_J = 
\begin{cases}
J , & J=1,2,
\\
\frac{J}{2}, & J \ge 3. 
\end{cases}
\eeq
}

\subsubsection*{Radial force} 
{
The radial component of the force per unit area on the shell is
\begin{equation} \label{F_abstr}
f(\theta) = \frac{\kappa}{a} \sum\limits_{j=1}^{J} ( x \cos\theta_j + y \sin\theta_j -w(\theta_j) ) \delta(\theta-\theta_j) ,
\end{equation} 
\rev{where $\delta(\theta)$ is the Dirac delta function.} The azimuthal component of the force is negligible. Substituting \eqref{0-5} into \eqref{F_abstr} yields 
\beq{094}
{
f(\theta) = -\frac{\kappa}{a} \Big( \frac{1}{\tau^2 - H_J} \Big) \sum\limits_{j=1}^{J} \bigg[ \sum\limits_{m=1}^{J} w(\theta_m) \cos(\theta_j-\theta_m) + \Big(\tau^2 - H_J\Big) w(\theta_j) \bigg] \delta(\theta-\theta_j) ,
}
\eeq
where $H_J$ is defined in \eqref{0-5}. The specific form of the radial force per unit length on the shell due to a single spring, a diametrical pair of springs, and for $J\ge 3$ uniformly distributed springs are given in eq.\ \eqref{8080}. 
 %as
%\begin{equation}
%f(\theta) = - \frac{\kappa}{a} \left(\frac{\tau^2}{\tau^2-1}\right)w(\theta_1) \delta(\theta-\theta_1) \qquad (J=1),
%\end{equation}
%and for  is
%\bal{f_two}
%f(\theta) = -\frac{\kappa}{a} \left( \frac{1}{\tau^2 - 2} \right) \bigg[ &\Big( (\tau^2-1)w(\theta_j) - w(\theta_j+\pi) \Big)  \delta(\theta-\theta_j) 
%\notag \\
%&+ \Big( - w(\theta_j) + (\tau^2-1)w(\theta_j+\pi) \Big) \delta(\theta-\theta_j-\pi)  \bigg] .
%\eal
%For $J\ge 3$  springs the total radial force per unit length on the shell is
%\beq{04943}
%f(\theta) = -\frac{\kappa}{a} \left( \frac{1}{\tau^2 - \frac{J}{2}} \right) \sum\limits_{j=1}^{J} \bigg[ \sum\limits_{m=1}^{J} w(\theta_m) \cos(\theta_j-\theta_m) + \Big(\tau^2 - \frac{J}{2}\Big) w(\theta_j) \bigg] \delta(\theta-\theta_j) .
%\eeq
}

{
\subsubsection*{Net force}
Expanding the radial force in equation \eqref{094} in azimuthal modes (see \eqref{-19}) for a single spring ($J=1$) at angle $\theta_1$ yields the modal force
\begin{equation}
f_n = - \frac{\kappa}{2\pi a} \left(\frac{\tau^2}{\tau^2 - 1} \right) w(\theta_1) \text{e}^{-\text{i}n\theta_1} .
\end{equation}
Similarly, for $J=2$ springs oriented at $\theta_1$ and $\theta_2 = \theta_1 +\pi$ the modal force has the form
\begin{subequations}
\begin{align}
f_n &= - \frac{\kappa}{2\pi a} \sum\limits_{j=1}^{2} w(\theta_j) \text{e}^{-in\theta_j}  \times
\begin{cases}
\frac{\tau^2}{\tau^2-2}  &  \ \text{for odd}\ n, 
\\
1 & \ \text{for even}\ n, 
\end{cases} 
\\
&= - \frac{\kappa}{2\pi a} \text{e}^{-\text{i}n\theta_1}  \times
\begin{cases}
\frac{\tau^2}{\tau^2-2} \left( w(\theta_1) - w(\theta_1+\pi) \right)  &  \ \text{for odd}\ n, 
\\
w(\theta_1) + w(\theta_1+\pi)  & \ \text{for even}\ n, 
\end{cases} 
\end{align}
\end{subequations}
Now consider the case $J\ge 3$, eq.\ \eqref{094} with $H_J=J/2$. In order to express the radial force as a Fourier series we first rewrite it as 
\beq{f_J}
f(\theta) = -\frac{\kappa}{2\pi a} \Big( \frac{1}{\tau^2 - \frac{J}{2}} \Big) \sum\limits_{n=-\infty}^{\infty}  \sum\limits_{j=1}^{J} w(\theta_j) \text{e}^{\text{i} n(\theta -\theta_j)} \bigg[ \tau^2 - \frac{J}{2}   +\sum\limits_{m=1}^{J} 
\cos(\theta_j-\theta_m) \text{e}^{\text{i} n(\theta_j-\theta_m)}  %\Big(
%\Big)  
\bigg]  .  
\eeq
Hence, referring to \eqref{-19}, 
\beq{838}
f_n = -\frac{\kappa}{2\pi a} \Big( \frac{1}{\tau^2 - \frac{J}{2}} \Big)  \sum\limits_{j=1}^{J} w(\theta_j) \text{e}^{-\text{i} n\theta_j} \bigg[  \tau^2 - \frac{J}{2}   + \sum\limits_{m=1}^{J} \cos(\theta_m-\theta_j) \text{e}^{-\text{i} n(\theta_m-\theta_j)}  \bigg] .
\eeq
Consider first the term 
\beq{3==7}
\sum\limits_{j=1}^{J} w(\theta_j) \text{e}^{\text{i}n\theta_j} 
= \sum\limits_{m=-\infty}^{\infty} W_m \sum\limits_{j=1}^{J} \text{e}^{\text{i}(m-n)\theta_j} 
= \sum\limits_{m=-\infty}^{\infty} W_m \hat{S}_{m-n}
, %\ \ \hat{z}=\text{e}^{i(m-n)\frac{2\pi}{J}} ,
\eeq
where 
\beq{0300}
\hat{S}_p=\sum\limits_{j=1}^{J} \text{e}^{\text{i}p\theta_j}  = \sum\limits_{j=1}^{J} \text{e}^{\text{i}j\theta_p} .
\eeq
For $p=0$\,mod\,$J$, $p\in \mathbb{Z}$, we have $\text{e}^{\text{i}\theta_p} = 1$  and 
hence $\hat{S}_p=J$. Otherwise
$\text{e}^{\text{i}\theta_p} \ne 1$ and therefore $\hat{S}_p=
(\text{e}^{\text{i} J\theta_p}-1)/(1-\text{e}^{-\text{i}\theta_p} )  = 0$. 
In conclusion,
\begin{equation} \label{w_proj}
\sum\limits_{j=1}^{J} w(\theta_j) \text{e}^{-\text{i}n\theta_j} =J \sum\limits_{m=-\infty}^{\infty} W_{n+mJ}.
\end{equation}
The modal force in \eqref{f_J} contains the summation
\beq{tr}
2 \sum\limits_{m=1}^{J} \cos(\theta_m-\theta_j) \text{e}^{-\text{i}n(\theta_m-\theta_j)} 
=2 \sum\limits_{m=1}^{J} \cos\theta_m \text{e}^{-\text{i}n\theta_m} 
=  \hat{S}_{1-n} + \hat{S}_{1+n}, 
\eeq
see \eqref{0300}. 
Thus 
\begin{equation} \label{trf}
 \sum\limits_{m=1}^{J} \cos(\theta_m-\theta_j) \text{e}^{-\text{i}n(\theta_m-\theta_j)}  = 
\begin{cases}
\frac J2, &  n= \pm 1\,\text{mod}\, J,  
\\
0, &  \text{otherwise},
\end{cases} 
\end{equation}
}
\rev{where the notation $n= \pm 1 \,\text{mod}\, J$ is defined in eq.~\eqref{nmod}.}
%Note that $z_{\pm} = -1$ for even $J$, but the even order polynomial in $z$ adds up to $S=0$.

Substituting results \eqref{w_proj} and \eqref{trf} into eq.\ \eqref{f_J} yields the modal force on the shell for $J\ge 3$ springs as 
\beq{-345}
f_n = -\frac{J \kappa}{2\pi a} \sum\limits_{m=-\infty}^{\infty} W_{n+mJ}  \times 
\begin{cases}
\frac{\tau^2}{\tau^2 - \frac{J}{2}} ,&   n= \pm 1\,\text{mod}\, J,  
\\
1 ,&  \text{otherwise}.
\end{cases}
\eeq
 
%%%%%%%%%%%%%%%%%%%%%%%%%%%%%%%%%%%%%%%%%%%%%%%%%%%%%%%%%%%%%%%%%%%%%%%%%%
%\bibliography{../../../SHARED_BIBLIOGRAPHY/AN_BIG_BIB}
%\bibliographystyle{unsrt}%natbib}%
%\end{document}

%%%%%%%%%%%%%%%%%%%%%%%%%%%%%%%%%%%%%%%%%%%%%%%%%%%%%%%%%%%%%%%%%%%%%%%%%%

\end{document}